\documentclass[useAMS,usenatbib,letterpaper]{mn2e}

\DeclareTextSymbol{\degre}{OT1}{23}
\usepackage{graphicx}
\usepackage{amsmath}
\usepackage{multirow}

\title[Accurate SB2 orbits]{Masses of the components of SB2 binaries observed with {\it Gaia}. IV. 
Accurate SB2 orbits for 14 binaries, and masses of 3 binaries
\footnotemark[1]\thanks{based on observations performed
at the Observatoire de Haute--Provence (CNRS), France}}
\author[F. Kiefer et al.]{F. Kiefer$^{1}$\thanks{E-mail:
flavien.kiefer@iap.fr}, 
J.-L. Halbwachs$^{2}$, Y. Lebreton$^{3,4}$, C. Soubiran$^{5}$, 
F. Arenou$^{3}$, D. Pourbaix$^{6}$,
\newauthor B. Famaey$^{2}$, P. Guillout$^{2}$,
R. Ibata$^{2}$ 
and T. Mazeh$^{7}$ \\
$^{1}$Institut d'Astrophysique de Paris,CNRS/UMR7095, 98bis boulevard Arago, 75014 Paris \\
$^{2}$Universit\'e de Strasbourg, CNRS, Observatoire astronomique de Strasbourg, UMR 7550, 
11 rue de l'Universit\'{e}, F-67000 Strasbourg, France\\
$^{3}$LESIA, Observatoire de Paris, PSL Research University, CNRS UMR 8109, Universit\'e Pierre et Marie Curie, Universit\'e Paris Diderot, \\ 5 Place Jules Janssen,
F-92195 Meudon, France\\
$^{4}$Institut de Physique de Rennes, Universit\'e de Rennes 1, CNRS UMR 6251, F-35042 Rennes, France\\
$^{5}$Laboratoire d'astrophysique de Bordeaux, Univ. Bordeaux, CNRS,  B18N, Allée Geoffroy Saint-Hilaire,  F-33615 PESSAC \\
$^{6}$FNRS, Institut d'Astronomie et d'Astrophysique, Universit\'{e} Libre de Bruxelles, boulevard du Triomphe, 1050 Bruxelles, Belgium\\
$^{7}$School of Physics and Astronomy, Tel Aviv University, Tel Aviv 69978, Israel\\
}

\begin{document}

\date{Accepted . Received 2017 ; in original form 2017}

\pagerange{\pageref{firstpage}--\pageref{lastpage}} \pubyear{2017}

\maketitle

\label{firstpage}

\begin{abstract} 
The orbital motion of non-contact double-lined spectroscopic binaries (SB2), with periods of a few tens of days to several years, holds unique accurate informations on individual stellar masses, that only long-term monitoring can unlock. The combination of radial velocity measurements from high-resolution spectrographs and astrometric measurements from high-precision interferometers allows the derivation of SB2 components masses down to the percent precision. Since 2010, we observed a large sample of SB2 with the SOPHIE spectrograph at the Observatoire de Haute-Provence, aiming at the derivation of orbital elements with sufficient accuracy to obtain masses of components with relative errors as low as 1\% when the astrometric measurements of the Gaia satellite will be taken into account. 
In this paper we present the results from six years of observations of 14 SB2 systems with periods ranging from 33 to 4185\,days. Using the {\sc todmor} algorithm we computed radial velocities from the spectra, and then derived the orbital elements of these binary systems. The minimum masses of the 28 stellar components are then obtained with a sample average accuracy of 1.0$\pm$0.2\,\%. Combining the radial velocities with existing interferometric measurements, we derived the masses of the primary and secondary components of HIP 61100, HIP\,95995 and HIP\,101382 with relative errors for components (A,B) of respectively (2.0, 1.7)\,\%, (3.7, 3.7)\,\%, and (0.2, 0.1)\,\%. Using the \verb+Cesam2k+ stellar evolution code, we could constrain the initial He-abundance, age and metallicity for HIP\,61100 and HIP\,95995. 
\end{abstract}

\begin{keywords}
binaries: spectroscopic, stars: fundamental parameters, 
stars: individual:HIP 61100, HIP 95995, HIP 101382
\end{keywords}


\section{Introduction}

Following the work of papers I-III~\citep{Halb2014,Halb2016,Kiefer2016} we propose to measure masses of stars with an accuracy better than 1\%. The loosely constrained single stars stellar evolution models still necessitate a confrontation to extremely accurate masses of stars. Non-contact binaries have the exclusive advantage to provide mass measurements of two separate objects with different masses but with the same age. They could thus provide a strong constraint on stellar models~\citep{Torres2012}. To that end, we proposed in paper I~\citep{Halb2014} to combine the high-resolution spectroscopy of the Spectrographe pour l'Observation des PH\'enom\`enes des Int\'erieurs Stellaires et des Exoplan\`etes (SOPHIE; Observatoire de Hautes-Provence) to the high-precision astrometry of the {\it Gaia} satellite on high-contrast large-period and bright spectroscopic binaries. SOPHIE provides 
radial velocities with an accuracy of a few tens of m~s$^{-1}$, and {\it Gaia} will soon deliver photocenter positions
with an accuracy of a few tens of microarcseconds. The combination of both will allow achieving better than 1\% accuracy on binary masses.

In paper I~\citep{Halb2014}, we selected a sample of 68 SB2s for which we expect to reach that level of precision. We have been observing these stars since 2010 with SOPHIE. A first result of our program was the detection of the secondary component in the spectra of 20 binaries which were previously known as single-lined (paper\,I). A second result was the determination of masses for two particular SB2s with accuracy between 0.26 and 2.4\,\%, coupling astrometric measurements from PIONIER and radial velocities from SOPHIE (paper\,II; Halbwachs et al. 2015). In a third paper (paper\,III; Kiefer et al. 2016), we derived projected masses (M $\sin^3i$) with precision better than 1.2\% for the two components of 10 binaries, and the masses of the binary HIP\,87895 with an accuracy of $\sim $1\% thanks to additional astrometric data.

Here, we present the accurate orbits measured for 14 SB2s (Table~\ref{tab:obs}) with periods ranging from 33 to 4185 days. After 8 years of observations with SOPHIE, we collected a total of 203 spectra of these stars. A large number of previously published measurements is also available for each of these binaries in the SB9 catalog~\citep{SB9}. Four of these targets were identified as new SB2s in paper I. We combined the radial velocity (RV) measurements and existing interferometric data 
for HIP\,61100, HIP\,95995 and HIP\,101382, to derive the masses of their components. 
This will enable us to validate the masses derived from our RVs and from Gaia astrometry,  
when the Gaia astrometric transits will be available.                                      
Meanwhile, in the present paper, these masses are confronted to evolutionary models. 

The observations are presented in Section~\ref{sect:observations}. The method of measurements of radial velocities from SOPHIE's observations is explained in Section~\ref{sect:RV}. We derive the orbital solutions in Section~\ref{sec:orbits}. The derivation of the masses of HIP\,61100, HIP\,95995 and HIP\,101382 is presented in Section~\ref{sec:masses}. Finally, in Section~\ref{sec:models} we examine how they compare to stellar evolution models.   


\section{Observations}
\label{sect:observations}

The observations were performed at the T193 telescope of the Haute-Provence Observatory, with the SOPHIE spectrograph~\citep{Perruchot2008}. SOPHIE is dedicated to the search of extrasolar planets, and, its high resolution ($R$$\sim $$75,000$) enables accurate stellar radial velocities to be measured for SB2 components. Since the beginning of the programme, we have accumulated 49 nights of observations in visitor mode. Before each observation run, ephemerides were derived from existing orbits provided by the SB9 catalogue \citep{SB9}, and priority classes were assigned on the basis of the orbital phase and of the observations already performed. In addition, we obtained observations in service mode for a total
duration of 7 nights; these observations were essentially used to complete the phase coverage of short-period binaries.

The spectra were all reduced through SOPHIE's pipeline, including localization of the orders on the frame, optimal order extraction, cosmic-ray rejection, wavelength calibration, flat-fielding and bias subtraction. 

Among all the observed SB2s, we have selected those which were satisfying two conditions:
\begin{itemize}
\item
They were observed over, at least, the part of the phase where the RVs of the
components may be derived. Except for HIP~77122, a binary with a period of more than 11 years, the observations
covered more than one period.
\item
They received a minimum of 11 observations. This limit was set for statistical reasons (see \textit{e.g.} paper III): Although an SB2 orbit could be derived in principle from only 6 of those observations, a minimum of 5 degrees-of-freedom on each component are needed for a reliable rescaling of the RV errorbars to the stochastic noise level,
as explained in Section~\ref{sec:orbits}. 
\end{itemize}
Table~\ref{tab:obs} summarizes this information.

\begin{table}
\caption{The SB2s analyzed in this paper.  }
\small
\begin{tabular}{@{}l@{~~}l@{~~}c@{~~}c@{~~}c@{~~}c@{~~}c@{~~}l@{}}
\hline
Name       & Alt. name  &  V       & Period$^a$ & $N_\text{spec}$ $^b$ & Span$^c$ & SNR$^d$  \\
HIP/HD     &  HD/BD            &(mag.)    &  (day)     &                      &   (period)    &       \\
\hline    
\multicolumn{7}{c}{\it Previously published SB2} \\
&&&&&& \\
HIP\,9121       & BD\,+41\,379  & 9.01    & 695       & 16              & 3.1          & 48    \\
HIP\,21946      & HD\,285970    & 9.86    &  56       & 11              & 34           & 54  \\
HIP\,38018      & HD\,61994     & 7.08    & 552       & 12              & 3.9          & 96   \\
HIP\,61100      & HD\,109011    & 8.10    & 1284      & 13              & 2.8          & 98   \\
HIP\,77122      & HD\,141335    & 8.95    & 4290      & 12              & 0.54         & 48   \\
HIP\,95995      & HD\,184467    & 6.59    &  494      & 14              & 4.3          & 145 \\
HIP\,100046     & HD\,193468    & 6.73    & 289       & 18              & 7.5          & 136 \\
HIP\,101382     & HD\,195987    & 7.09    & 57        & 18              & 45           & 101 \\
HIP\,116360     & HD\,221757    & 7.22    & 348       & 15              & 9.8          & 97 \\
HD\,98031       & BD\,+13\,2380 & 8.40    & 271       & 15              & 7.9          & 48  \\
\hline
\multicolumn{7}{c}{\it SB2s identified in paper I, previously published as SB1s} \\
&&&&&& \\
HIP\,7143       & HD\,9312      & 6.81    & 37        & 16              & 59           & 143    \\
HIP\,12472      & HD\,16646     & 8.10    & 329       & 13              & 6.7          & 90    \\
HIP\,48895      & HD\,86358     & 6.46    & 34        & 17              & 87           & 140    \\
HIP\,72706      & HD\,131208    & 7.61    & 84        & 13              & 18           & 97   \\
\hline
\end{tabular}\\
\flushleft
$^a$ The period values are taken from the SB9 catalogue~\citep{SB9}. Except HIP\,3818 and HIP\,61100~\citep{Halb2003}. \\
$^b$ $ N_\text{spec}$ gives the number of spectra collected with SOPHIE. \\
$^c$ Span is the total time span of the observation epochs used in the orbit derivation, counted in number of periods. \\
$^d$ SNR is the median signal-to-noise ratio of all the SOPHIE spectra of a given star at 5550\,\AA.
\label{tab:obs}
\end{table}


\section{Radial velocity measurements}
\label{sect:RV}

The radial velocities of the components are derived using the TwO-Dimensional cross-CORrelation algorithm {\sc todcor} \citep{zucker94,zucker04}. 
It calculates the cross-correlation of an SB2 spectrum and two best-matching stellar atmosphere models, one for each component of the observed binary system. The radial velocities of both components are measured at the optimum of this two-dimensional cross-correlation function (CCF). 
More specifically we employed the multi-order version of {\sc todcor} that is named {\sc todmor}~\citep{zucker04}. We redirect the readers to our preceding articles (paper I-III) for more details on this algorithm. 

All SOPHIE multi-orders spectra were first corrected for the blaze using the response function provided by SOPHIE's pipeline; then for each of them, the pseudo-continuum was detrended using a p-percentile filter~\citep[paper II,][]{Hodg1985}. Finally, before deriving the RVs, a best-matching model spectrum is determined for both SB2 components of each star.

\subsection{Optimizing the model spectra}
The theoretical spectra from the PHOENIX stellar atmosphere models~\citep{Huss2013} optimized for best-matching of the components of all 14 SB2s are given in Table~\ref{tab:stellpar}.
Contrary to the method presented in previous papers, instead of optimizing the CCF for all orders, 
we minimized the $\chi^2$ of the selected spectra compared to the PHOENIX models around the Ca\,I line at $\sim$6120\,\AA (order 33). This line is particularly sensitive to $T_\text{eff}$ and $\log g$ if conditions are close to LTE~\citep{Drake1991,Mashonkina2007}. Moreover being on the red side of the spectrum it offers the best conditions with respect to signal-to-noise and strength of the second component. We used the full order $\#$ 33, which also incorporates a few Fe lines. 
Compared to the previous method explained in paper III, which consisted in optimizing the CCF, the above method led to more reliable values of stellar parameters, with in particular less biased values of metallicity. We verified that the two methods give consistent, and equally satisfying, results on radial velocities measurements. 

We optimized the values of effective temperature $T_\text{eff}$, rotational broadening $v \sin i$, metallicity [Fe/H], surface gravity $\log(g)$, and flux ratio at $4916$\,\AA, $\alpha$$=$$F_2/F_1$. For binaries on the main sequence, if $\alpha$ is too low ($\alpha$$<$$0.1$) and the secondary $\log g$ cannot be properly derived, we fixed its value with respect to $T_\text{eff}$, following the empirical relation $\log g = 12 - 2\log T_\text{eff}$, as derived from Fig. 1 of~\cite{Angelov1996}. Each theoretical spectrum is convolved with the instrument line spread function, here modeled by a Gaussian, and pseudo-continuum detrended with the same techniques employed for detrending the observed spectra. 

The values of the stellar parameters, and their uncertainties, given in Table~\ref{tab:stellpar} are the average and standard deviation of the individual estimations. The 1$\sigma$ uncertainties do not include known systematics, see e.g.~\cite{Torres2012}. To give a point of comparison, we measured the Sun's parameters on SOPHIE spectra of Vesta and Ceres in Table~\ref{tab:stellpar}. Metallicity was found to be off by $-0.12$ dex, $\Delta T_{\text{eff},\odot}$$\sim $$60$\,K, and $\Delta\log(g)_\odot$$\sim $$0.1$. Given their small amplitude, and lacking an exhaustive analysis of benchmark stars spectra with this method, these errors could be considered as more realistic minimum uncertainties on $T_\text{eff}$, $\log(g)$ and [Fe/H], than the values given in Table~\ref{tab:stellpar}. Furthermore, the uncertainty on the effective temperatures is actually larger than this, since varying by hand metallicity in a $\pm$0.1\,dex range for a few targets, we found an amplitude of variations of $T_\text{eff}$ on the order of 100-200\,K.

\newcommand{\MS}{\text{\tiny MS}}
\begin{table}
\begin{minipage}{\columnwidth}
\caption{\label{tab:stellpar}The stellar parameters of the 14 SB2, determined by $\chi^2$ optimization around the Ca\,I line at 6121\,\AA. Sun's parameters derived with the same protocol are given in the last row.}
\scriptsize
\begin{tabular}{@{}lrrrrr@{}}
\hline 
HIP         & $^a$\,\!$ T_\text{eff,1}$ &  $^b$\,\!$ \log g_1$   &  $^c$\,\!$ V_1 \sin i_1$  &  $^d$\,\!$ [\text{Fe/H}]$ &  $\alpha$   \\ 
HD          & $ T_\text{eff,2}$ &  $ \log g_2$   &  $ V_2 \sin i_2$&                  &                \\ 
            & (K)               &  (dex)         &   (km s$^{-1}$)   &     (dex)        & (flux ratio)   \\ 
\hline 
HIP\,7143   & 5367        & 4.30        & 4.7       & 0.03        & 0.063        \\ 
            & $\pm$166    & $\pm$0.13   & $\pm$0.6  & $\pm$0.03   & $\pm$0.013  \\ 
            & 5150        & 4.6$_\MS$   & $<$8      &             &              \\ 
            & $\pm$228    &             &           &             &              \\  \\
HIP\,9121   & 5789        & 4.40        & 2.7       & 0.11        & 0.083        \\ 
            & $\pm$21     & $\pm$0.03   & $\pm$0.3  & $\pm$0.01   & $\pm$0.010   \\ 
            & 4544        & 4.89        & $<$2      &             &              \\ 
            & $\pm$164    & $\pm$0.10   &           &             &              \\ \\
HIP\,12472  & 6253        & 4.4$_\MS$   & 10.5      & -0.86       & 0.037        \\ 
            & $\pm$82     &             & $\pm$0.3  & $\pm$0.12   & $\pm$0.012  \\ 
            & 4802        & 4.6$_\MS$   & $<$4      &             &              \\ 
            & $\pm$292    &             &           &             &              \\ \\
HIP\,21946  & 4680        & 4.72        & 4.0       & -0.13       & 0.035        \\ 
            & $\pm$21     & $\pm$0.05   & $\pm$0.4  & $\pm$0.03   & $\pm$0.004   \\ 
            & 4164        & 4.8$_\MS$   & $<$10     &             &              \\ 
            & $\pm$110    &             &           &             &              \\  \\
HIP\,38018  & 5585        & 4.46        & 3.5       & -0.04       & 0.069        \\ 
            & $\pm$13     & $\pm$0.04   & $\pm$0.2  & $\pm$0.06   & $\pm$0.011   \\ 
            & 4484        & 4.7$_\MS$   & $<$5      &             &              \\ 
            & $\pm$110    &             &           &             &              \\ \\
HIP\,48895  & 6186        & 4.30        & 74.1      & -0.59       & 0.253        \\ 
            & $\pm$152    & $\pm$0.09   & $\pm$2.5  & $\pm$0.08   & $\pm$0.010  \\ 
            & 5697        & 4.69        & 21.4      &             &              \\ 
            & $\pm$79     & $\pm$0.10   & $\pm$1.0  &             &              \\  \\

HIP\,61100  & 5105        & 4.75        & 5.7       & -0.14       & 0.229        \\ 
            & $\pm$21     & $\pm$0.10   & $\pm$0.1  & $\pm$0.10   & $\pm$0.004   \\ 
            & 4175        & 4.8         & 5.1       &             &              \\ 
            & $\pm$35     & $\pm$0.1    & $\pm$0.4  &             &              \\  \\
HIP\,72706  & 4524        & 3.27        & 4.3       & -0.13       & 0.099        \\ 
            & $\pm$8      & $\pm$0.03   & $\pm$0.4  & $\pm$0.01   & $\pm$0.023  \\ 
            & 5272        & 4.50        & 2.9       &             &              \\ 
            & $\pm$280    & $\pm$0.11   & $\pm$1.1  &             &              \\ \\
HIP\,77122  & 5638        & 4.22        & $<$4      & -1          & 0.195        \\ 
            & $\pm$45     & $\pm$0.24   &           &             & $\pm$0.060  \\ 
            & 5035        & 4.60        & $<$3      &             &              \\ 
            & $\pm$131    & $\pm$0.31   &           &             &              \\ \\
HIP\,95995  & 5114        & 4.62        & 2.7       & -0.33       & 0.524        \\ 
            & $\pm$11     & $\pm$0.05   & $\pm$0.3  & $\pm$0.01   & $\pm$0.083  \\ 
            & 4705        & 4.67        & 2.7       &             &              \\ 
            & $\pm$101    & $\pm$0.05   & $\pm$1.0  &             &              \\  \\
HD\,98031   & 6018        & 4.55        & 2.4       & -0.13       & 0.236        \\ 
            & $\pm$8      & $\pm$0.07   & $\pm$0.7  & $\pm$0.04   & $\pm$0.001  \\ 
            & 5095        & 4.86        & $<$3      &             &              \\ 
            & $\pm$19     & $\pm$0.07   &           &             &              \\  \\
HIP\,100046 & 6069        & 4.28        & 26.2      & -0.71       & 0.585        \\ 
            & $\pm$53     & $\pm$0.21   & $\pm$1.0  & $\pm$0.06   & $\pm$0.016  \\ 
            & 5623        & 4.36        & 13.7      &             &              \\ 
            & $\pm$43     & $\pm$0.21   & $\pm$0.5  &             &              \\  \\
HIP\,101382 & 5296        & 4.71        & 3.9       & -0.38       & 0.156        \\ 
            & $\pm$19     & $\pm$0.03   & $\pm$0.3  & $\pm$0.01   & $\pm$0.005  \\ 
            & 4360        & 4.97        & $<$2      &             &              \\ 
            & $\pm$87     & $\pm$0.04   &           &             &              \\  \\
HIP\,116360 & 6227        & 4.37        & 4.3       & -0.21       & 0.624        \\ 
            & $\pm$68     & $\pm$0.09   & $\pm$0.4  & $\pm$0.02   & $\pm$0.026   \\ 
            & 5915        & 4.49        & 3.5       &             &              \\ 
            & $\pm$10     & $\pm$0.09   & $\pm$1.1  &             &              \\  \\
Sun         & 5836        & 4.58        & 4.9       & -0.12       &        \\ 
            & $\pm$40     & $\pm$0.10   & $\pm$0.2  & $\pm$0.04   &   \\ 
\hline
\end{tabular}
\flushleft
$^a$Minimum systematic uncertainties on T$_\text{eff}$ are about 100\,K. \\
$^b$The MS subscript indicates that the $\log g$ did not converge to a realistic value ($>5$) and was fixed to be on the Main Sequence following $\log g = 12 - 2\log T_\text{eff}$~\citep{Angelov1996}. \\
$^c$When the $V \sin i$ is compatible with zero, we give the upper bound at the 1$\sigma$ limit; $V\sin i$=0 is used to derive the RVs. \\
$^d$Given the systematic error on [Fe/H]$_\text{sun}$, a more reliable value of uncertainty on [Fe/H] should be at least 0.1\,dex.
\end{minipage} 
\end{table}

\subsection{Deriving radial velocities}
We then applied {\sc todmor} to all multi-order spectra of each target and determined the radial velocities of both components discarding all orders harboring strong telluric lines, following the method of paper III. 

In the cases where the S/N ratio and the secondary-to-primary flux ratio were large enough (SNR$>$90 and $\alpha$$>$$0.1$), we incorporated an enhancement on the 2D-CCF calculation. We employed the first-derivative of the spectra, rather than the spectra themselves. Using first derivative is equivalent to applying a linear filter on the spectra, filtering out low frequency components (like \textit{e.g.} the continuum). Unfortunately, it enhances high frequency noise, and for that reason cannot be used on low S/N ratio spectra. We found that it greatly reduces systematics on RV measurements of those binaries with strong blend. 

Final velocities for each component are displayed in Table~\ref{tab:RVs}. They are used to derive the orbital solutions for the 14 SB2s in the next section. 

\begin{table*} 
\caption{\label{tab:RVs} New radial velocities from SOPHIE and obtained with {\sc todmor}. The uncertainties must still be corrected
as explained in Section~\ref{sec:orbits}.
Outliers are marked with an asterisk ($^*$) and are not taken into account in the analysis.}
\scriptsize
\begin{minipage}{89mm}
\begin{tabular}{@{}l@{~~}c@{~~}c@{~~}c@{~~}c@{~~}c@{~~}c@{}} 
\hline  
\multicolumn{7}{c}{HIP 7143} \\ 
&&&&&&   \\ 
BJD      & $RV_1$        & $\sigma_{RV 1}$ & $RV_2$        & $\sigma_{RV 2}$ & $O_1-C_1$   & $O_2-C_2$ \\ 
-2400000 & km s$^{-1}$   & km s$^{-1}$      & km s$^{-1}$   & km s$^{-1}$      & km s$^{-1}$ & km s$^{-1}$      \\ 
\hline  
2455440.5949    &     -28.6137      &   0.0087            &    39.5047       &   0.0869    & -0.0112 & -0.2463       \\ 
2455532.3039    &     29.9337      &   0.0088            &    -36.9342       &   0.0868    & -0.0103 & 0.0246       \\ 
2455783.6041    &     17.3891      &   0.0087            &    -20.3808       &   0.0806    & -0.0093 & 0.1403       \\ 
2455864.4055    &     30.4795      &   0.0089            &    -37.6110       &   0.0869    & 0.0215 & 0.0213       \\ 
2456148.5899    &     16.4093      &   0.0087            &    -19.2077       &   0.0891    & -0.0247 & 0.0498       \\ 
2456243.3400    &     -25.1332      &   0.0087            &    35.2395       &   0.0865    & 0.0346 & -0.0113       \\ 
2456323.2855    &     -32.9010      &   0.0091            &    45.6786       &   0.0881    & -0.0284 & 0.3327       \\ 
2456525.5388$^*$    &     -6.8752$^*$      &   0.0087$^*$            &    -26.1124$^*$       &   0.1173$^*$    & -29.9263$^*$ & 1.8151$^*$       \\ 
2456525.5489    &     23.0257      &   0.0087            &    -28.0582       &   0.0882    & 0.0047 & -0.1701       \\ 
2456526.5759    &     19.7230      &   0.0087            &    -23.4145       &   0.0856    & 0.0054 & 0.1453       \\ 
2456619.4717    &     -7.1678      &   0.0088            &    11.3499       &   0.0840    & -0.0101 & -0.3033       \\ 
2456889.5967    &     26.0816      &   0.0089            &    -31.9416       &   0.0877    & 0.0086 & -0.0546       \\ 
2457009.3357    &     -20.1976      &   0.0088            &    28.8898       &   0.0874    & 0.0199 & 0.1252       \\ 
2457414.3022    &     -35.4380      &   0.0109            &    48.4965       &   0.1051    & 0.0110 & -0.2250       \\ 
2457602.5998    &     -26.8715      &   0.0087            &    37.4743       &   0.0842    & -0.0264 & 0.0259       \\ 
2457635.5346    &     -38.7686      &   0.0087            &    53.2214       &   0.0843    & 0.0135 & 0.1327       \\ 
&&&&&&   \\ 
&&&&&&   \\ 
\end{tabular}

\end{minipage}%
\begin{minipage}{89mm}
\begin{tabular}{@{}l@{~~}c@{~~}c@{~~}c@{~~}c@{~~}c@{~~}c@{}} 
\hline  
\multicolumn{7}{c}{HIP 9121} \\ 
&&&&&&   \\ 
BJD      & $RV_1$        & $\sigma_{RV 1}$ & $RV_2$        & $\sigma_{RV 2}$ & $O_1-C_1$   & $O_2-C_2$ \\ 
-2400000 & km s$^{-1}$   & km s$^{-1}$      & km s$^{-1}$   & km s$^{-1}$      & km s$^{-1}$ & km s$^{-1}$      \\ 
\hline  
2455440.6065    &     -4.4125      &   0.0097            &    15.9387       &   0.0857    & -0.0110 & 0.3651       \\ 
2455532.3126    &     20.1786      &   0.0097            &    -17.2062       &   0.0868    & -0.0138 & 0.0057       \\ 
2455605.3068    &     18.1418      &   0.0109            &    -14.5813       &   0.0939    & -0.0412 & -0.0481       \\ 
2455864.4267    &     0.1595      &   0.0102            &    8.4879       &   0.0950    & -0.0357 & -0.9579       \\ 
2456147.5908    &     -3.7162      &   0.0089            &    15.0394       &   0.0794    & 0.0164 & 0.3576       \\ 
2456243.3503    &     24.8212      &   0.0098            &    -23.5053       &   0.0845    & 0.0156 & -0.1438       \\ 
2456323.3008    &     14.9387      &   0.0099            &    -10.3337       &   0.0958    & 0.0112 & -0.1404       \\ 
2456525.6005    &     1.4133      &   0.0100            &    7.5680       &   0.0781    & -0.0094 & -0.2414       \\ 
2456618.4926    &     -1.6987      &   0.0099            &    11.8527       &   0.0976    & 0.0128 & -0.1348       \\ 
2456913.4916    &     16.2085      &   0.0099            &    -12.0895       &   0.0840    & 0.0032 & -0.1928       \\ 
2456919.4382    &     19.2128      &   0.0106            &    -15.7803       &   0.0949    & -0.0096 & 0.1384       \\ 
2457009.3471    &     16.0572      &   0.0093            &    -11.9517       &   0.0823    & 0.0135 & -0.2704       \\ 
2457073.2958    &     9.4184      &   0.0103            &    -1.5753       &   0.1178    & 0.0239 & 1.2421       \\ 
2457295.6294    &     -1.1549      &   0.0095            &    11.0601       &   0.0913    & 0.0282 & -0.2231       \\ 
2457414.3261    &     -4.2774      &   0.0176            &    15.4326       &   0.1592    & -0.0186 & 0.0493       \\ 
2457603.5503    &     13.7650      &   0.0098            &    -8.4294       &   0.0887    & -0.0101 & 0.2277       \\ 
&&&&&&   \\ 
&&&&&&   \\ 
\end{tabular}

\end{minipage}\\%
\begin{minipage}{89mm}
\begin{tabular}{@{}l@{~~}c@{~~}c@{~~}c@{~~}c@{~~}c@{~~}c@{}} 
\hline  
\multicolumn{7}{c}{HIP 12472} \\ 
&&&&&&   \\ 
BJD      & $RV_1$        & $\sigma_{RV 1}$ & $RV_2$        & $\sigma_{RV 2}$ & $O_1-C_1$   & $O_2-C_2$ \\ 
-2400000 & km s$^{-1}$   & km s$^{-1}$      & km s$^{-1}$   & km s$^{-1}$      & km s$^{-1}$ & km s$^{-1}$      \\ 
\hline  
2455532.4082    &     1.6055      &   0.0441            &    -14.5375       &   0.0810    & -0.0746 & 0.4260       \\ 
2455605.3484    &     6.6888      &   0.0809            &    -22.8412       &   0.1327    & 0.0108 & 0.0023       \\ 
2455783.6130    &     -13.7153      &   0.0469            &    9.2047       &   0.0741    & -0.0131 & -0.0843       \\ 
2455864.5337    &     2.5696      &   0.0443            &    -16.3092       &   0.0803    & 0.0459 & -0.0156       \\ 
2455933.2717    &     6.7945      &   0.0901            &    -23.0360       &   0.1543    & -0.0399 & 0.0543       \\ 
2456148.6166$^*$    &     -8.1223$^*$      &   0.0412$^*$           &    2.5239$^*$       &   0.0800$^*$    & 0.1609$^*$ & 1.7787$^*$       \\ 
2456243.3940    &     9.0360      &   0.0451            &    -26.7895       &   0.0804    & 0.0032 & -0.2332       \\ 
2456525.5750    &     3.4028      &   0.0455            &    -18.0037       &   0.0779    & 0.0256 & -0.3644       \\ 
2456618.5243    &     0.6876      &   0.0438            &    -12.7734       &   0.0642    & 0.0213 & 0.5918       \\ 
2456889.6184    &     9.0899      &   0.0443            &    -26.7878       &   0.0744    & 0.0123 & -0.1609       \\ 
2457009.3760    &     -11.9803      &   0.0427            &    6.2414       &   0.0842    & -0.0057 & -0.3238       \\ 
2457603.5777    &     0.9406      &   0.0407            &    -12.9910       &   0.0671    & -0.0580 & 0.8981       \\ 
2457721.5573    &     -15.4946      &   0.0431            &    12.2045       &   0.0737    & -0.0072 & 0.1008       \\ 
&&&&&&   \\ 
&&&&&&   \\ 
&&&&&&   \\ 
&&&&&&   \\ 
&&&&&&   \\ 
\end{tabular}

\end{minipage}%
\begin{minipage}{89mm}
\begin{tabular}{@{}l@{~~}c@{~~}c@{~~}c@{~~}c@{~~}c@{~~}c@{}} 
\hline  
\multicolumn{7}{c}{HIP 21946} \\ 
&&&&&&   \\ 
BJD      & $RV_1$        & $\sigma_{RV 1}$ & $RV_2$        & $\sigma_{RV 2}$ & $O_1-C_1$   & $O_2-C_2$ \\ 
-2400000 & km s$^{-1}$   & km s$^{-1}$      & km s$^{-1}$   & km s$^{-1}$      & km s$^{-1}$ & km s$^{-1}$      \\ 
\hline  
2455864.5931    &     30.9873      &   0.0119            &    -9.4994       &   0.1903    & 0.0049 & -0.0450       \\ 
2456243.5337    &     22.6048      &   0.0129            &    2.1410       &   0.2291    & 0.0031 & -0.2630       \\ 
2456323.4334    &     26.0735      &   0.0130            &    -2.6027       &   0.1837    & -0.0138 & -0.0747       \\ 
2456619.5802    &     -5.5830      &   0.0129            &    42.5907       &   0.2249    & -0.0007 & 0.3067       \\ 
2457009.4399    &     11.3879      &   0.0126            &    18.2601       &   0.1442    & -0.0030 & -0.0070       \\ 
2457014.5018    &     -4.8917      &   0.0128            &    41.3349       &   0.2290    & -0.0147 & 0.0489       \\ 
2457020.4023    &     -21.7157      &   0.0124            &    64.5834       &   0.2309    & 0.0133 & -0.5478       \\ 
2457073.3398    &     -14.1075      &   0.0134            &    55.1816       &   0.2816    & -0.0206 & 0.8638       \\ 
2457295.6463    &     -0.8778      &   0.0125            &    35.4673       &   0.2188    & 0.0024 & -0.1632       \\ 
2457699.5165    &     -18.9958      &   0.0149            &    60.8245       &   0.2310    & -0.0132 & -0.4207       \\ 
2457761.3441    &     4.5465      &   0.0131            &    28.4629       &   0.2285    & 0.0065 & 0.5018       \\ 
&&&&&&   \\ 
&&&&&&   \\ 
&&&&&&   \\ 
&&&&&&   \\ 
&&&&&&   \\ 
&&&&&&   \\ 
&&&&&&   \\ 
\end{tabular}

\end{minipage}\\%
\begin{minipage}{89mm}
\begin{tabular}{@{}l@{~~}c@{~~}c@{~~}c@{~~}c@{~~}c@{~~}c@{}} 
\hline  
\multicolumn{7}{c}{HIP 38018} \\ 
&&&&&&   \\ 
BJD      & $RV_1$        & $\sigma_{RV 1}$ & $RV_2$        & $\sigma_{RV 2}$ & $O_1-C_1$   & $O_2-C_2$ \\ 
-2400000 & km s$^{-1}$   & km s$^{-1}$      & km s$^{-1}$   & km s$^{-1}$      & km s$^{-1}$ & km s$^{-1}$      \\ 
\hline  
2455605.4554    &     -15.8212      &   0.0080            &    -31.6573       &   0.0610    & -0.0658 & -0.0322       \\ 
2455966.3859    &     -29.6282      &   0.0076            &    -10.8293       &   0.0617    & 0.0218 & -0.0811       \\ 
2456034.3241    &     -36.0742      &   0.0115            &    -1.1332       &   0.1010    & -0.0046 & -0.0305       \\ 
2456243.6090    &     -15.3134      &   0.0076            &    -32.0804       &   0.0613    & 0.0189 & 0.1804       \\ 
2456324.3742    &     -17.3822      &   0.0075            &    -29.3241       &   0.0780    & 0.0105 & -0.1591       \\ 
2456619.6080    &     -31.0369      &   0.0075            &    -8.7959       &   0.0653    & 0.0025 & -0.1352       \\ 
2456700.4676    &     -16.2770      &   0.0084            &    -31.0950       &   0.0668    & 0.0377 & -0.3103       \\ 
2457009.5244    &     -23.9774      &   0.0072            &    -18.7759       &   0.0526    & -0.0100 & 0.5106       \\ 
2457073.3764    &     -29.7104      &   0.0062            &    -10.6737       &   0.0518    & 0.0035 & -0.0214       \\ 
2457159.3587    &     -34.2367      &   0.0087            &    -3.9517       &   0.0743    & -0.0118 & -0.0772       \\ 
2457728.6301    &     -30.3165      &   0.0072            &    -9.8280       &   0.0596    & -0.0067 & -0.0711       \\ 
2457734.6292    &     -28.5850      &   0.0070            &    -12.1420       &   0.0588    & -0.0027 & 0.2105       \\ 
&&&&&&   \\ 
&&&&&&   \\ 
&&&&&&   \\ 
&&&&&&   \\ 
&&&&&&   \\ 
&&&&&&   \\ 
\end{tabular}

\end{minipage}%
\begin{minipage}{89mm}
\begin{tabular}{@{}l@{~~}c@{~~}c@{~~}c@{~~}c@{~~}c@{~~}c@{}} 
\hline  
\multicolumn{7}{c}{HIP 48895} \\ 
&&&&&&   \\ 
BJD      & $RV_1$        & $\sigma_{RV 1}$ & $RV_2$        & $\sigma_{RV 2}$ & $O_1-C_1$   & $O_2-C_2$ \\ 
-2400000 & km s$^{-1}$   & km s$^{-1}$      & km s$^{-1}$   & km s$^{-1}$      & km s$^{-1}$ & km s$^{-1}$      \\ 
\hline  
2455966.4971    &     41.4163      &   0.1379            &    27.6481       &   0.0794    & -2.7001 & 3.7556       \\ 
2456243.6398    &     51.9407      &   0.1321            &    5.7655       &   0.0789    & -0.3817 & -2.0201       \\ 
2456323.5312    &     26.0378      &   0.1738            &    59.6220       &   0.0749    & 1.6606 & -3.0149       \\ 
2456414.3350    &     51.5298      &   0.1408            &    13.2361       &   0.0804    & 2.6904 & -1.3861       \\ 
2456619.6215    &     46.9870      &   0.2076            &    26.1839       &   0.0769    & 5.5179 & -2.9049       \\ 
2456701.4324    &     25.4852      &   0.1369            &    63.5982       &   0.0748    & 0.4116 & 2.3281       \\ 
2456764.3623    &     22.6437      &   0.1384            &    67.3399       &   0.0753    & 0.6953 & -0.0644       \\ 
2457009.5859    &     35.7275      &   0.1334            &    40.4026       &   0.0801    & -2.0947 & 4.1557       \\ 
2457073.4027    &     26.4163      &   0.1506            &    59.2057       &   0.0701    & -0.9983 & 2.5305       \\ 
2457160.3757    &     37.5641      &   0.1408            &    32.6651       &   0.0771    & -0.2168 & -3.6628       \\ 
2457505.3652    &     20.7102      &   0.1568            &    66.4075       &   0.0792    & -1.5991 & -0.2883       \\ 
2457759.6114    &     53.2215      &   0.1481            &    5.0854       &   0.0833    & 0.0073 & -0.9498       \\ 
2457792.3706    &     52.3538      &   0.1550            &    6.4234       &   0.0860    & -1.0702 & 0.8000       \\ 
2454845.6013    &     23.3993      &   0.1296            &    66.5367       &   0.0702    & 0.6445 & 0.7153       \\ 
2454846.6471    &     23.7001      &   0.1339            &    65.2222       &   0.0700    & -0.3565 & 1.9559       \\ 
2454847.6167    &     24.5010      &   0.1368            &    61.5615       &   0.0692    & -1.2396 & 1.6006       \\ 
2454848.6314    &     23.7993      &   0.1569            &    58.3759       &   0.0726    & -4.1640 & 2.7777       \\ 
&&&&&&   \\ 
\end{tabular}

\end{minipage}
\vspace*{1cm}
\end{table*}

\addtocounter{table}{-1}
\begin{table*}
\caption{Continued.}
\scriptsize
\begin{minipage}{89mm}
\begin{tabular}{@{}l@{~~}c@{~~}c@{~~}c@{~~}c@{~~}c@{~~}c@{}} 
\hline  
\multicolumn{7}{c}{HIP 98031} \\ 
&&&&&&   \\ 
BJD      & $RV_1$        & $\sigma_{RV 1}$ & $RV_2$        & $\sigma_{RV 2}$ & $O_1-C_1$   & $O_2-C_2$ \\ 
-2400000 & km s$^{-1}$   & km s$^{-1}$      & km s$^{-1}$   & km s$^{-1}$      & km s$^{-1}$ & km s$^{-1}$      \\ 
\hline  
2455692.3492    &     72.8755      &   0.0109            &    64.6313       &   0.0348    & 0.0058 & 0.2213       \\ 
2455933.6379    &     74.1453      &   0.0179            &    62.9888       &   0.0565    & -0.0012 & 0.0166       \\ 
2455966.5107    &     72.7083      &   0.0107            &    64.8335       &   0.0326    & 0.0266 & 0.2119       \\ 
2456324.4501    &     62.7036      &   0.0129            &    75.7316       &   0.0386    & 0.0783 & -0.2146       \\ 
2456414.3490    &     70.9941      &   0.0109            &    66.2017       &   0.0307    & -0.0852 & -0.2245       \\ 
2456619.6559    &     60.5365      &   0.0104            &    78.3963       &   0.0308    & 0.0011 & 0.0967       \\ 
2456700.5552    &     72.8220      &   0.0106            &    64.6617       &   0.0330    & -0.0164 & 0.2165       \\ 
2456763.4017    &     73.6712      &   0.0104            &    63.6086       &   0.0334    & 0.0479 & 0.0473       \\ 
2457009.6140    &     74.2814      &   0.0104            &    62.6262       &   0.0324    & 0.0175 & -0.2138       \\ 
2457159.3958    &     60.6343      &   0.0136            &    78.2756       &   0.0403    & 0.0247 & 0.0597       \\ 
2457160.3675    &     60.5873      &   0.0109            &    78.3338       &   0.0325    & 0.0103 & 0.0811       \\ 
2457436.6010    &     60.5133      &   0.0104            &    78.4542       &   0.0308    & -0.0138 & 0.1452       \\ 
2457471.5766    &     65.6334      &   0.0133            &    72.2121       &   0.0384    & -0.0192 & -0.3250       \\ 
2457505.3820    &     71.8385      &   0.0108            &    65.6050       &   0.0305    & -0.0450 & 0.0844       \\ 
2457819.5101    &     74.2800      &   0.0102            &    62.6705       &   0.0316    & 0.0109 & -0.1636       \\ 
&&&&&&   \\ 
&&&&&&   \\ 
&&&&&&   \\ 
\end{tabular}

\end{minipage}%
\begin{minipage}{89mm}
\begin{tabular}{@{}l@{~~}c@{~~}c@{~~}c@{~~}c@{~~}c@{~~}c@{}} 
\hline  
\multicolumn{7}{c}{HIP 61100} \\ 
&&&&&&   \\ 
BJD      & $RV_1$        & $\sigma_{RV 1}$ & $RV_2$        & $\sigma_{RV 2}$ & $O_1-C_1$   & $O_2-C_2$ \\ 
-2400000 & km s$^{-1}$   & km s$^{-1}$      & km s$^{-1}$   & km s$^{-1}$      & km s$^{-1}$ & km s$^{-1}$      \\ 
\hline  
2454124.7212    &     -2.1884      &   0.0123            &    -19.3714       &   0.0703    & 0.0204 & 0.0195       \\ 
2454889.6065$^*$    &     -14.1818$^*$      &   0.0147$^*$            &    -4.7802$^*$       &   0.0840$^*$    & -0.0628$^*$ & -0.9102$^*$       \\ 
2455306.4091    &     -4.0880      &   0.0110            &    -16.6121       &   0.0709    & 0.0729 & 0.2349       \\ 
2455605.5531    &     -4.1704      &   0.0095            &    -16.9148       &   0.0623    & -0.0524 & -0.0118       \\ 
2456243.6803    &     -16.1346      &   0.0096            &    -1.3190       &   0.0642    & -0.0225 & -0.0463       \\ 
2456323.5971    &     -18.8578      &   0.0095            &    2.4018       &   0.0570    & -0.0414 & 0.1503       \\ 
2456413.3839    &     -21.4210      &   0.0121            &    5.7921       &   0.0727    & -0.0032 & 0.1504       \\ 
2456619.6953    &     -2.9770      &   0.0095            &    -18.4678       &   0.0569    & 0.0052 & -0.0846       \\ 
2456700.5934    &     -2.2159      &   0.0096            &    -19.3730       &   0.0534    & 0.0059 & 0.0011       \\ 
2456763.5299    &     -2.6145      &   0.0109            &    -18.9232       &   0.0639    & 0.0107 & -0.0748       \\ 
2457505.4205    &     -15.4310      &   0.0093            &    -2.6677       &   0.0635    & 0.0064 & -0.5160       \\ 
2457759.6959    &     -18.4149      &   0.0092            &    1.8460       &   0.0559    & -0.0069 & 0.1266       \\ 
2457767.6521    &     -17.3820      &   0.0088            &    0.4519       &   0.0530    & 0.0041 & 0.0643       \\ 
&&&&&&   \\ 
&&&&&&   \\ 
&&&&&&   \\ 
&&&&&&   \\ 
&&&&&&   \\ 
\end{tabular}

\end{minipage}\\%
\begin{minipage}{89mm}
\begin{tabular}{@{}l@{~~}c@{~~}c@{~~}c@{~~}c@{~~}c@{~~}c@{}} 
\hline  
\multicolumn{7}{c}{HIP 72706} \\ 
&&&&&&   \\ 
BJD      & $RV_1$        & $\sigma_{RV 1}$ & $RV_2$        & $\sigma_{RV 2}$ & $O_1-C_1$   & $O_2-C_2$ \\ 
-2400000 & km s$^{-1}$   & km s$^{-1}$      & km s$^{-1}$   & km s$^{-1}$      & km s$^{-1}$ & km s$^{-1}$      \\ 
\hline  
2456033.5020    &     -61.2850      &   0.0075            &    -25.7871       &   0.1070    & -0.0192 & 0.1112       \\ 
2456147.3536    &     -19.4972      &   0.0073            &    -79.3190       &   0.1022    & 0.0474 & 0.1668       \\ 
2456324.5831    &     -26.9623      &   0.0074            &    -70.0314       &   0.0976    & -0.0366 & -0.0260       \\ 
2456414.4587    &     -32.8521      &   0.0073            &    -62.4272       &   0.1031    & -0.0029 & -0.0302       \\ 
2456700.6800    &     -60.4920      &   0.0073            &    -27.0406       &   0.1136    & -0.0494 & -0.0850       \\ 
2457073.6330    &     -24.3145      &   0.0067            &    -73.3307       &   0.0989    & -0.0167 & 0.0499       \\ 
2457159.4832    &     -26.5475      &   0.0072            &    -70.5603       &   0.0951    & -0.0049 & -0.0631       \\ 
2457470.6254    &     -68.1682      &   0.0082            &    -16.7236       &   0.1115    & 0.0102 & 0.2961       \\ 
2457475.6035$^*$    &     -49.3856$^*$      &   0.0154$^*$            &    -43.8833$^*$       &   0.1826$^*$    & 0.0671$^*$ & -2.8120$^*$       \\ 
2457505.5470    &     -37.3135      &   0.0072            &    -56.6873       &   0.1064    & -0.0023 & -0.0213       \\ 
2457542.3640    &     -65.4462      &   0.0146            &    -21.0450       &   0.2545    & -0.0316 & -0.4755       \\ 
2457542.4266    &     -65.4616      &   0.0077            &    -20.7041       &   0.1106    & 0.0008 & -0.1960       \\ 
2457550.5131    &     -70.0817      &   0.0076            &    -14.6032       &   0.1050    & 0.0109 & -0.0421       \\ 
2457819.5824    &     -19.7787      &   0.0072            &    -79.2520       &   0.1005    & -0.0295 & -0.0292       \\ 
&&&&&&   \\ 
&&&&&&   \\ 
&&&&&&   \\ 
&&&&&&   \\ 
\end{tabular}

\end{minipage}%
\begin{minipage}{89mm}
\begin{tabular}{@{}l@{~~}c@{~~}c@{~~}c@{~~}c@{~~}c@{~~}c@{}} 
\hline  
\multicolumn{7}{c}{HIP 77122} \\ 
&&&&&&   \\ 
BJD      & $RV_1$        & $\sigma_{RV 1}$ & $RV_2$        & $\sigma_{RV 2}$ & $O_1-C_1$   & $O_2-C_2$ \\ 
-2400000 & km s$^{-1}$   & km s$^{-1}$      & km s$^{-1}$   & km s$^{-1}$      & km s$^{-1}$ & km s$^{-1}$      \\ 
\hline  
2455306.5079$^*$    &     -49.1723$^*$      &   0.0097$^*$            &    -44.2639$^*$       &   0.0300$^*$    & -0.4494$^*$ & -0.5203$^*$       \\ 
2456033.5194    &     -53.1250      &   0.0125            &    -38.9871       &   0.0349    & 0.0055 & -0.2407       \\ 
2456148.3858    &     -54.7882      &   0.0124            &    -37.0043       &   0.0349    & 0.0060 & -0.1444       \\ 
2456324.6472    &     -60.8632      &   0.0117            &    -30.0466       &   0.0327    & -0.0091 & -0.0572       \\ 
2456413.5714    &     -78.1954      &   0.0120            &    -10.2543       &   0.0345    & 0.0058 & 0.0674       \\ 
2456414.4769    &     -78.2975      &   0.0119            &    -10.1390       &   0.0331    & -0.0017 & 0.0754       \\ 
2456525.3381    &     -36.6294      &   0.0119            &    -57.4821       &   0.0337    & 0.0001 & -0.0273       \\ 
2456764.5216    &     -39.7897      &   0.0115            &    -53.8866       &   0.0328    & -0.0018 & -0.0126       \\ 
2457073.6595    &     -41.6506      &   0.0108            &    -51.5984       &   0.0312    & 0.0179 & 0.1433       \\ 
2457159.4975    &     -42.0369      &   0.0112            &    -51.2538       &   0.0320    & 0.0088 & 0.0604       \\ 
2457505.5611    &     -43.2999      &   0.0115            &    -49.8530       &   0.0337    & -0.0202 & 0.0621       \\ 
2457602.3944    &     -43.5835      &   0.0118            &    -49.5196       &   0.0359    & -0.0104 & 0.0628       \\ 
&&&&&&   \\ 
&&&&&&   \\ 
&&&&&&   \\ 
&&&&&&   \\ 
&&&&&&   \\ 
&&&&&&   \\ 
\end{tabular}

\end{minipage}\\%
\begin{minipage}{89mm}
\begin{tabular}{@{}l@{~~}c@{~~}c@{~~}c@{~~}c@{~~}c@{~~}c@{}} 
\hline  
\multicolumn{7}{c}{HIP 95995} \\ 
&&&&&&   \\ 
BJD      & $RV_1$        & $\sigma_{RV 1}$ & $RV_2$        & $\sigma_{RV 2}$ & $O_1-C_1$   & $O_2-C_2$ \\ 
-2400000 & km s$^{-1}$   & km s$^{-1}$      & km s$^{-1}$   & km s$^{-1}$      & km s$^{-1}$ & km s$^{-1}$      \\ 
\hline  
2455440.3925    &     14.1543      &   0.0130            &    9.6101       &   0.0166    & -0.0109 & -0.0654       \\ 
2455693.5801    &     14.9946      &   0.0122            &    8.8627       &   0.0158    & -0.0041 & 0.0420       \\ 
2455784.4255    &     17.5962      &   0.0115            &    6.1163       &   0.0152    & -0.0029 & -0.0379       \\ 
2456034.6088    &     0.4528      &   0.0115            &    23.7456       &   0.0151    & -0.0055 & 0.0139       \\ 
2456243.2705    &     17.0740      &   0.0116            &    6.7064       &   0.0153    & 0.0032 & 0.0105       \\ 
2456414.5806    &     15.0522      &   0.0122            &    8.8434       &   0.0159    & 0.0010 & 0.0765       \\ 
2456525.3813    &     1.0363      &   0.0114            &    23.1434       &   0.0150    & 0.0034 & 0.0009       \\ 
2456618.3725    &     8.9973      &   0.0123            &    14.8330       &   0.0160    & 0.0057 & -0.1480       \\ 
2456890.4648    &     15.9419      &   0.0118            &    7.9785       &   0.0157    & 0.0063 & 0.1185       \\ 
2457295.3224    &     17.6864      &   0.0114            &    6.0110       &   0.0151    & 0.0012 & -0.0548       \\ 
2457295.3224    &     17.6864      &   0.0114            &    6.0110       &   0.0151    & 0.0012 & -0.0548       \\ 
2457505.6066    &     2.5417      &   0.0113            &    21.5832       &   0.0148    & -0.0038 & -0.0081       \\ 
2457525.5532    &     -0.6032      &   0.0115            &    24.8503       &   0.0149    & -0.0002 & 0.0303       \\ 
2457556.4625    &     0.1786      &   0.0114            &    24.0395       &   0.0147    & -0.0016 & 0.0225       \\ 
&&&&&&   \\ 
&&&&&&   \\ 
&&&&&&   \\ 
&&&&&&   \\ 
\end{tabular}

\end{minipage}%
\begin{minipage}{89mm}
\begin{tabular}{@{}l@{~~}c@{~~}c@{~~}c@{~~}c@{~~}c@{~~}c@{}} 
\hline  
\multicolumn{7}{c}{HIP 100046} \\ 
&&&&&&   \\ 
BJD      & $RV_1$        & $\sigma_{RV 1}$ & $RV_2$        & $\sigma_{RV 2}$ & $O_1-C_1$   & $O_2-C_2$ \\ 
-2400000 & km s$^{-1}$   & km s$^{-1}$      & km s$^{-1}$   & km s$^{-1}$      & km s$^{-1}$ & km s$^{-1}$      \\ 
\hline  
2455440.3983    &     -2.7679      &   0.0647            &    -32.9372       &   0.0372    & -1.1721 & -0.1793       \\ 
2455693.5912    &     -10.4280      &   0.0563            &    -24.0270       &   0.0345    & -0.8709 & 0.0598       \\ 
2455864.2976    &     -30.1345      &   0.0683            &    -0.9387       &   0.0389    & 0.7532 & -0.0845       \\ 
2456034.6180    &     2.4505      &   0.0598            &    -37.1059       &   0.0326    & -0.0083 & 0.0681       \\ 
2456147.4411    &     -31.6547      &   0.0562            &    0.5238       &   0.0309    & 0.5029 & -0.0051       \\ 
2456414.5964    &     -37.1051      &   0.0567            &    5.3736       &   0.0328    & -0.4000 & -0.1083       \\ 
2456525.4010    &     -16.3540      &   0.0549            &    -16.7767       &   0.0341    & -0.1247 & 0.0429       \\ 
2456619.3099    &     4.4352      &   0.0994            &    -38.9722       &   0.0569    & 0.3452 & -0.0215       \\ 
2456890.4961    &     -1.6800      &   0.0535            &    -33.6484       &   0.0312    & -0.7823 & -0.1301       \\ 
2456940.3394    &     6.2028      &   0.0551            &    -40.5074       &   0.0332    & 0.5430 & 0.1531       \\ 
2457159.5328    &     -6.9649      &   0.0548            &    -28.4290       &   0.0332    & -1.2451 & -0.1627       \\ 
2457295.3591    &     -34.1782      &   0.0474            &    2.6393       &   0.0291    & 0.0230 & -0.1154       \\ 
2457505.6238    &     9.5644      &   0.0521            &    -43.9000       &   0.0317    & 0.9413 & -0.0121       \\ 
2457511.5904    &     9.7700      &   0.0514            &    -44.1377       &   0.0300    & 0.8712 & 0.0506       \\ 
2457539.5993    &     -30.5416      &   0.0526            &    -0.3724       &   0.0314    & 0.4782 & 0.3380       \\ 
2457542.4749    &     -34.3064      &   0.0557            &    2.8589       &   0.0337    & -0.1168 & 0.1167       \\ 
2457563.5620    &     -38.9690      &   0.0490            &    7.1852       &   0.0301    & -0.7511 & 0.0556       \\ 
2457602.4760    &     -29.7396      &   0.0512            &    -1.1213       &   0.0315    & 0.8755 & 0.0298       \\ 
\end{tabular}

\end{minipage}
\vspace*{1cm}
\end{table*}

\addtocounter{table}{-1}
\begin{table*}
\caption{Continued.}
\scriptsize
\begin{minipage}{89mm}
\begin{tabular}{@{}l@{~~}c@{~~}c@{~~}c@{~~}c@{~~}c@{~~}c@{}} 
\hline  
\multicolumn{7}{c}{HIP 101382} \\ 
&&&&&&   \\ 
BJD      & $RV_1$        & $\sigma_{RV 1}$ & $RV_2$        & $\sigma_{RV 2}$ & $O_1-C_1$   & $O_2-C_2$ \\ 
-2400000 & km s$^{-1}$   & km s$^{-1}$      & km s$^{-1}$   & km s$^{-1}$      & km s$^{-1}$ & km s$^{-1}$      \\ 
\hline  
2455306.6279$^*$    &     27.4215$^*$      &   0.0056$^*$            &    -48.4736$^*$       &   0.0279$^*$    & -0.5988$^*$ & -0.7284$^*$       \\ 
2455440.4752    &     -16.0573      &   0.0065            &    8.1802       &   0.0339    & -0.0550 & -0.0675       \\ 
2456147.4603    &     -23.0348      &   0.0069            &    17.0744       &   0.0350    & 0.0072 & -0.1270       \\ 
2456243.2796    &     -16.4840      &   0.0068            &    9.0005       &   0.0344    & 0.0097 & 0.1278       \\ 
2456525.4373    &     -7.2323      &   0.0064            &    -2.7826       &   0.0362    & 0.0153 & 0.1050       \\ 
2456889.4781$^*$    &     -24.9314$^*$      &   0.0604$^*$            &    18.7325$^*$       &   0.3251$^*$    & -0.0833$^*$ & -0.7662$^*$       \\ 
2456890.5346    &     -24.3951      &   0.0065            &    18.8520       &   0.0332    & -0.0102 & -0.0576       \\ 
2456906.5610    &     7.4644      &   0.0066            &    -21.5996       &   0.0337    & -0.0098 & 0.0128       \\ 
2456914.4368    &     32.1885      &   0.0062            &    -53.0538       &   0.0316    & -0.0126 & 0.0088       \\ 
2456922.2759    &     6.8794      &   0.0065            &    -20.9364       &   0.0328    & -0.0106 & -0.0670       \\ 
2457159.5648    &     -15.0283      &   0.0065            &    7.0422       &   0.0331    & 0.0041 & 0.0282       \\ 
2457160.5553    &     -16.6925      &   0.0065            &    9.1756       &   0.0331    & 0.0054 & 0.0433       \\ 
2457295.3722    &     -21.7360      &   0.0058            &    15.6240       &   0.0282    & 0.0032 & 0.0795       \\ 
2457568.5135    &     -23.8142      &   0.0066            &    18.2184       &   0.0341    & 0.0142 & 0.0166       \\ 
2457571.5894    &     -25.1569      &   0.0091            &    19.9035       &   0.0466    & 0.0011 & 0.0106       \\ 
2457602.4876    &     32.1200      &   0.0066            &    -52.9594       &   0.0333    & -0.0002 & 0.0005       \\ 
2457606.3739    &     22.0613      &   0.0122            &    -40.1714       &   0.0607    & 0.0035 & -0.0100       \\ 
2457883.6019    &     18.4379      &   0.0065            &    -35.5057       &   0.0317    & 0.0264 & 0.0179       \\ 
\end{tabular}

\end{minipage}%
\begin{minipage}{89mm}
\begin{tabular}{@{}l@{~~}c@{~~}c@{~~}c@{~~}c@{~~}c@{~~}c@{}} 
\hline  
\multicolumn{7}{c}{HIP 116360} \\ 
&&&&&&   \\ 
BJD      & $RV_1$        & $\sigma_{RV 1}$ & $RV_2$        & $\sigma_{RV 2}$ & $O_1-C_1$   & $O_2-C_2$ \\ 
-2400000 & km s$^{-1}$   & km s$^{-1}$      & km s$^{-1}$   & km s$^{-1}$      & km s$^{-1}$ & km s$^{-1}$      \\ 
\hline  
2454339.5767    &     15.9868      &   0.0096            &    37.7172       &   0.0119    & -0.0488 & -0.0609       \\ 
2454345.4505    &     15.7252      &   0.0107            &    38.0853       &   0.0130    & -0.0111 & -0.0142       \\ 
2454408.3317    &     15.4871      &   0.0096            &    38.3689       &   0.0118    & -0.0028 & 0.0047       \\ 
2454852.2873    &     31.2565      &   0.0198            &    21.3992       &   0.0241    & -0.0501 & 0.0261       \\ 
2455784.5553    &     15.0810      &   0.0119            &    38.8450       &   0.0147    & 0.0180 & 0.0222       \\ 
2456147.5212    &     15.4694      &   0.0122            &    38.4331       &   0.0150    & 0.0177 & 0.0278       \\ 
2456525.4971    &     17.2681      &   0.0116            &    36.4516       &   0.0140    & 0.0056 & -0.0084       \\ 
2456618.4151    &     45.4494      &   0.0117            &    6.1845       &   0.0144    & -0.0116 & 0.0170       \\ 
2456889.5524    &     18.9561      &   0.0111            &    34.6675       &   0.0136    & -0.0164 & 0.0445       \\ 
2456962.3075    &     42.7728      &   0.0077            &    9.0399       &   0.0095    & -0.0214 & 0.0075       \\ 
2456990.3484    &     55.6703      &   0.0113            &    -4.7991       &   0.0138    & 0.00001 & 0.0007       \\ 
2457295.4256    &     34.3595      &   0.0097            &    18.0903       &   0.0119    & -0.0182 & 0.0164       \\ 
2457295.4256    &     34.3595      &   0.0097            &    18.0903       &   0.0119    & -0.0182 & 0.0164       \\ 
2457603.5447    &     21.8455      &   0.0108            &    31.5873       &   0.0131    & 0.0589 & -0.0127       \\ 
2457732.2988    &     32.4490      &   0.0109            &    20.1658       &   0.0134    & 0.0062 & 0.0133       \\ 
&&&&&&   \\ 
&&&&&&   \\ 
&&&&&&   \\ 
\end{tabular}

\end{minipage}
\vspace*{1cm}
\end{table*}


\section{The SB2 orbits}
\label{sec:orbits}

The orbits derived from the RVs in Table~\ref{tab:RVs} have too large residuals in relation to uncertainties.
This is clear when the $F_2$ indicator of the goodness-of-fit (GOF) is calculated (see paper II, equation 1): the values are 
too large to obey a normal distribution, because the uncertainties were underestimated. This results in underestimating
the uncertainties of the parameters of the orbit, but also in assigning erroneous weights to the RVs of each component.
Deriving an SB2 orbital solution necessitates attributing realistic errors to each dataset properly. The correction
process  was already used in paper III, we refer the reader to that paper for explanations. The corrected errors express as follow with respect to initial errorbars $\sigma_{RV}$:

\begin{align}
\sigma^\text{corr}_{RV, 1} &= \varphi_1 \times \sqrt{\sigma_{RV, 1}^2 + \varepsilon_1^2} \label{eq:correction1}\\
\sigma^\text{corr}_{RV, 2} &= \varphi_2 \times \sqrt{\sigma_{RV, 2}^2 + \varepsilon_2^2} \label{eq:correction2}
\end{align}

The correction terms $\varphi_1$, $\varphi_2$, $\varepsilon_1$ and $\varepsilon_2$ are given in Table~\ref{tab:corsigRVprev}. 
The references of previously published RVs are also displayed in this table, as well as the related correction terms.

The orbital solutions of the 14 SB2s were derived twice: from the SOPHIE RVs alone, and also combining them with 
the previously published RVs. The results are presented on Table~\ref{tab:orbSB2}. 

Only the  period, $P$, the time of periastron passage,
$T_0$, and the SOPHIE offset $d_{n-p}$ were taken from the combined solution, since $P$ and $T_0$ are more accurate than in the SOPHIE solution.
The eccentricity $e$, the center-of-mass velocity $\gamma$,
the periastron longitude $\omega$, the RV amplitudes $K_1$ and $K_2$ and the deducted 
minimum masses and minimum semi-major axes were all taken from the SOPHIE solution. The primary offset $d_{2-1}$
also refers to this solution. The secondary component velocities are often shifted by up to a few 100\,m s$^{-1}$ compared primary's velocities (paper II-III). This $d_{2-1}$ incorporates such shift as an additional parameter to the RV fit.

\begin{table*}
\caption{Correction terms applied to the uncertainties of the previous and of the new RV measurements. The composition of these terms into a uncertainty correction is set out in Section~\ref{sec:orbits}, eqs.~\ref{eq:correction1} and~\ref{eq:correction2}. When the original publication provides only weights for the previous measurements,
$\varphi_{1,p}$ and $\varphi_{2,p}$ are the uncertainties corresponding to $W=1$, for the primary and for the secondary component, respectively.
}
\begin{tabular}{llcccccccc}
\hline
HIP/HD   & Reference of previous RV  &  \multicolumn{4}{c}{Correction terms for previous measurements} &  \multicolumn{4}{c}{Correction terms for new measurements}   \\
       &                        &$\varepsilon_{1,p}$&$\varphi_{1,p}$&$\varepsilon_{1,p}$&$\varphi_{2,p}$&$\varepsilon_{1,n}$&$\varphi_{1,n}$& $\varepsilon_{2,n}$& $\varphi_{2,n}$\\
       &                        &km s$^{-1}$        &               & km s$^{-1}$       &               & km s$^{-1}$       &               &  km s$^{-1}$       &          \\
\hline
HIP 7143   & \cite{Katoh13}         &       0.017       & 1             & $\ldots$           & $\ldots$       &  0.0129           & 0.947         & 0.1768             & 0.947    \\
HIP 9121   & \cite{Goldberg02}      &       0           & 0.599         & 0                 & 3.962         & 0.0232            & 0.937         & 0.5146             & 0.937 \\
HIP 12472  & CGG95$^a$              &       0           & 0.674         & $\ldots$           & $\ldots$       & 0                 &  1.049        & 0.2515             &  1.049  \\
HIP 21946  & HMU12$^a$              & 0                 & 1             & 0                 & 1             & 0.0068            &  1.116        & 0.2676             &  1.116 \\
HIP 38018  & DM88$^a$               & 0                 & 1             & $\ldots^b$         & $\ldots^b$   & 0.0338            &  0.921        & 0.2519             &  0.921\\
HIP 48895  & \cite{Griffin06}$^c$   & $\ldots$           & $\ldots$       & 0                 & 1.409         & 1.4918            & 1.054         &  0.6993            & 1.054  \\
HD 98031  & \cite{Griffin05b}       & 0                 & 0.316         & 0                 & 0.822         & 0.0233            & 0.961         & 0.2224             & 0.961 \\
HIP 61100  & \cite{Halb2003}     & 0                 & 0.7359        & 1.467             & 1             & 0.0196            & 1.021         & 0.1965             &  1.021 \\
HIP 72706  & \cite{Massarotti08}    & 0                 & 1             & $\ldots$           & $\ldots$       & 0.0267            & 1.208         & 0                  &  1.208 \\
HIP 77122  & \cite{Goldberg02}$^d$  & 0.770             & 1             & 3.416             & 1             &  0.0085           & 1.083         &  0.0993            & 1.083 \\
HIP 95995  & \cite{Pourbaix00}      &       0           & 1.070         & 0                 & 1.060         &  0                & 0.489         & 0.0736             & 0.953   \\
HIP 100046  & \cite{Griffin05c}$^e$ & 0                 & 0.471         & 0                 & 0.404         & 0.7031            & 1.060         & 0.0905             & 1.060 \\
HIP 101382  & \cite{Torres2002}       & 0                 & 0.316         & 0                 & 1.422         & 0.0206            & 0.938         & 0.0658             & 0.938 \\
HIP 116360  & \cite{Griffin05a}     & 0                 & 0.325         & 0                 & 0.433         & 0.0272            &  0.982        & 0.0238             &  0.982\\
\hline
\label{tab:corsigRVprev}
\end{tabular}
\flushleft 
$^a$ CGG95 = \cite{Carquillat95}, HMU12 = \cite{Halb2012}, DM88 = \cite{Duquennoy88}\\
$^b$ The secondary component was marginally detected by DM88, but these measurements were so inaccurate that we prefer to ignore them.\\
$^c$ \cite{Griffin06} derived the orbit of the secondary component, but not of the primary.
He detected the primary component four times, but he didn't take these measurements into account in the
derivation of the orbit.\\
$^d$ We have discarded the primary RV obtained by \cite{Goldberg02} for the epoch BJD=$2\,446\,604.810$.\\
$^e$ The components of \cite{Griffin05b} are in the reverse position.
\end{table*}

\begin{table*}
 \centering
 \begin{minipage}{178mm}
  \caption{The orbital elements of the 14 SB2s. Apart for HIP~77122, $P$ and $T_0$ were derived from the previously published RV measurements and from the new ones, but the
other elements correspond to the new RVs alone. The radial velocity of the barycentre, $V_0$, is in the reference 
system of the new measurements of the primary component. 
The minimum masses and minimum semi-major axes are derived from the true period 
($P_{true}=P \times (1-V_0/c)$). The numbers in parentheses refer to the previously published RV measurements}
\scriptsize
  \begin{tabular}{@{}lrrrrrrrrrrrrrr@{}}
  \hline
HIP & $P$ & $T_0$(BJD) & $e$ &  $V_0$ & $\omega_1$ & $K_1$ &  ${\cal M}_1 \sin^3 i$ &  $a_1 \sin i$&$N_1$&   $d_{n-p}$ & $\sigma(O_1-C_1)$ \\
HD/BD  &     &           &     &        &            & $K_2$ &  ${\cal M}_2 \sin^3 i$ &  $a_2 \sin i$&$N_2$& $d_{2-1}$ &$\sigma(O_2-C_2)$    \\
    & (d) & 2400000+  &  &(km s$^{-1}$)&($^{\rm o}$)&(km s$^{-1}$)&(${\cal M}_\odot$)&  (Gm) & & (km s$^{-1}$) &(km s$^{-1}$)    \\
  \hline
HIP 7143   & 36.519182   & 56614.6542 & 0.14321      & 0.7726     & 203.418   & 34.9715    &1.0971     & 17.3808    & 15&  -2.7542   &0.012\\
HD 9312   &$\pm 0.000031$&$\pm 0.0057$&$\pm 0.00018$&$\pm 0.0040$&$\pm 0.073$&$\pm 0.0057$&$\pm0.0033$&$\pm 0.0028$ & (12)   & $\pm 0.0085$ &(0.019)      \\
       &              &            &              &            &           & 45.821      &0.8373     &22.773     &  15& 0.490     &0.173\\
       &              &            &              &            &           &$\pm 0.065$  &$\pm0.0014$&$\pm 0.032$&    &$\pm 0.049$&     \\
&&&&&&&&&&&\\ 
HIP 9121    & 694.613   & 56921.910  & 0.56841      & 4.1302     & 313.220  & 15.108    &1.003      & 118.725   & 16& 0.338   &0.020\\
BD +41 379 &$\pm 0.022$&$\pm 0.069$ &$\pm 0.00054$&$\pm 0.0068$&$\pm 0.081$&$\pm 0.012$&$\pm0.020$ &$\pm 0.082$&(40)& $\pm 0.097$  &(0.517)\\
       &             &            &              &            &       & 20.14     &0.7524     &158.3      & 16& 0.070     &0.452\\
       &             &            &              &            &       &$\pm 0.19$ &$\pm0.0080$&$\pm 1.5$  &(39)&$\pm 0.135$& (3.52)    \\
&&&&&&&&&&&\\ 
HIP 12472   & 328.800   & 56891.43 & 0.1451     & -4.929     & 354.28  & 12.341    &0.6478      & 55.19    & 11& 0.076  &0.031\\
HD 16646   &$\pm 0.020$&$\pm 0.72$&$\pm 0.0021$&$\pm 0.021$&$\pm 0.84$&$\pm 0.022$&$\pm0.0089$ &$\pm 0.10$&(67)& $\pm 0.102$ & (0.861)\\
        &           &         &           &            &          & 19.44     &0.4112     & 86.94     & 11& 0.383     &0.274\\
       &            &            &       &            &       &$\pm 0.12$   &$\pm0.0034$&$\pm 0.53$  &   &$\pm 0.102$&     \\
&&&&&&&&&&&\\ 
HIP 21946   & 56.44365    & 56964.6923 & 0.35377     & 14.2337    & 191.221   & 26.7000    &0.7515     & 19.3823    & 11&  0.427  &0.010\\
HD 285970  &$\pm 0.00016$&$\pm 0.0079$&$\pm 0.00024$&$\pm 0.0064$&$\pm 0.067$&$\pm 0.0084$&$\pm0.0077$&$\pm 0.0066$&(38)&  $\pm 0.081$  &(0.404)\\
        &             &            &             &            &           & 37.78      &0.5312     & 27.42      & 11& 0.011     &0.406\\
        &             &            &             &            &           &$\pm 0.18$  &$\pm0.0030$&$\pm 0.13$  &(3)&$\pm 0.135$&(5.17)\\
&&&&&&&&&&&\\ 
HIP 38018   & 553.206    & 57163.64& 0.4276     & -22.194    & 222.21   & 10.549    &0.4676     & 72.54    & 12&   0.400   &0.024\\
HD 61994   &$\pm 0.037$ &$\pm 0.19$&$\pm 0.0012$&$\pm 0.011$ &$\pm 0.22$&$\pm 0.015$&$\pm0.0067$&$\pm 0.11$&(17)&  $\pm 0.115$ &(1.26)\\
        &             &        &            &            &           & 15.85    &0.3112     &108.98    & 12& 0.243     &0.215\\
        &             &        &            &            &           &$\pm 0.10$&$\pm0.0026$&$\pm 0.70$&   &$\pm 0.078$&     \\
&&&&&&&&&&&\\ 
HIP 48895 & 33.71218 &56574.65  &0.0565 & 37.02  &310.2  &15.82  &0.2373 &  7.32  & 17   & -1.845 & 1.569      \\
HD 86358 &$\pm 0.00091$&$\pm 0.77$&$\pm 0.0087$&$\pm  0.40$&$\pm 8.9$&$\pm 0.51$ &$\pm0.0070$&$\pm  0.24$&      &  $\pm  0.381$ &            \\
      &          &          &       &        &       &31.06  &0.1209 & 14.37  & 17   & 0.776       & 0.576      \\
      &          &          &       &        &       &$\pm 0.26$&$\pm 0.0067$&$\pm 0.12$& (32)  &$\pm 0.465$&(1.67)      \\
&&&&&&&&&&&\\ 
HD 98031& 271.265    & 56637.63 & 0.2216     & 68.664    & 213.63   & 6.8751     &0.04309     & 24.996    & 14&  -0.650   &0.020\\
BD +13 2380&$\pm 0.017$ &$\pm 0.37$&$\pm 0.0019$&$\pm 0.014$&$\pm 0.63$&$\pm 0.0088$&$\pm0.00083$&$\pm 0.033$&(65)&  $\pm 0.050$ &(0.301)\\
        &             &        &            &            &           & 7.742     &0.03827     &28.15    & 14& 0.482     &0.202\\
        &             &        &            &            &           &$\pm 0.072$&$\pm0.00039$&$\pm 0.26$&(65)&$\pm 0.064$&(1.13)\\
&&&&&&&&&&&\\ 
HIP 61100   & 1284.11    & 56493.58& 0.5119     & -9.722    & 244.75   & 9.615    &0.5186     & 145.94    & 12&  0.343         &0.016\\
HD 109011  &$\pm 0.14$   &$\pm 0.37$&$\pm 0.0012$&$\pm 0.011$&$\pm 0.20$&$\pm 0.013$&$\pm0.0069$&$\pm 0.16$&(35)& $\pm 0.065$  &(0.361)     \\
        &             &        &            &            &           & 12.530    &0.3980     &190.2    & 12& 0.122     &0.203\\
        &             &        &            &            &           &$\pm 0.078$&$\pm0.0029$&$\pm 1.2$&(35)&$\pm 0.065$&(2.44) \\
&&&&&&&&&&&\\ 
HIP 72706   & 83.52955    & 56474.385& 0.49100    &-46.056   &  275.75    & 25.308    &0.6218     & 25.328    & 12&  0.107 &0.022\\
HD 131208  &$\pm 0.00085$&$\pm 0.036$&$\pm 0.00058$&$\pm 0.023$&$\pm 0.14$&$\pm 0.019$&$\pm0.0020$&$\pm 0.014$&(16)& $\pm 0.091$ &(0.388)\\
        &             &        &            &            &        & 32.506    &0.48411     &32.532    & 12& 0.622     &0.123\\
        &             &        &            &            &        &$\pm 0.050$&$\pm0.00099$&$\pm 0.048$&   &$\pm 0.062$&    \\
&&&&&&&&&&&\\ 
HIP 77122$^a$& 4185.42     & 56423.36& 0.94077     &-46.5890    &  239.00    & 21.363    &0.8507     & 416.92    & 11&  0.111   &0.011\\
HD 141335  &$\pm 0.55$   &$\pm 0.20$&$\pm 0.00021$&$\pm 0.0061$&$\pm 0.11$  &$\pm 0.018$&$\pm0.0047$&$\pm 0.48$ &(54)& $\pm 0.108$ &(0.816)\\
        &             &          &             &            &            & 24.221    &0.7503     & 472.7       & 11& 0.426     &0.112\\
        &             &          &             &            &            &$\pm 0.055$&$\pm0.0031$&$\pm 1.1$  &(55)&$\pm 0.038$&(3.42)\\
&&&&&&&&&&&\\ 
HIP 95995   & 494.313     & 56549.487& 0.38926     &11.8933     &  180.300   & 9.4911     &0.14399     & 59.424    & 13&  0.592      &0.0045\\
HD 184467  &$\pm 0.012$ &$\pm 0.046$&$\pm 0.00030$&$\pm 0.0018$&$\pm 0.043$ &$\pm 0.0026$&$\pm0.00076$&$\pm 0.015$&(36)& $\pm 0.106$ &(0.596)\\
        &             &          &             &            &            & 9.733      &0.14041     &60.94      & 13& 0.112     &0.067\\
        &             &          &             &            &            &$\pm 0.026$ &$\pm0.00038$&$\pm 0.16$ &(36)&$\pm 0.021$&(0.860)\\
&&&&&&&&&&&\\ 
HIP 100046  & 289.4669    & 56661.48 & 0.5704     &-16.52    &  83.28    & 23.90     &1.078     & 78.13    & 18&  -1.410    &0.750\\
HD 193468  &$\pm 0.0076$ &$\pm 0.13$&$\pm 0.0013$&$\pm 0.18$&$\pm 0.24$ &$\pm 0.24$ &$\pm0.011$&$\pm 0.79$&(70)&  $\pm 0.311$ &(1.57)\\
        &             &          &             &         &           & 26.031    &0.990     &85.11     & 18& 0.017     &0.079\\
        &             &          &             &         &           &$\pm 0.040$&$\pm0.020$&$\pm 0.11$&(70)&$\pm 0.188$&(0.809)\\
&&&&&&&&&&&\\ 
HIP 101382  & 57.32176    & 56627.3786 & 0.30514     &-5.4108     &  357.195  & 28.8519    &0.8088      & 21.6578    & 15&  0.411      &0.017\\
HD 195987  &$\pm 0.00010$&$\pm 0.0075$&$\pm 0.00025$&$\pm 0.0057$&$\pm 0.064$&$\pm 0.0077$&$\pm0.0012$ &$\pm 0.0056$&(73)& $\pm 0.038$   &(0.322)\\
        &             &            &             &            &           & 36.697     &0.63590    &27.547      & 15& 0.187     &0.063\\
        &             &            &             &            &           &$\pm 0.025$ &$\pm0.00058$ &$\pm 0.019$ &(73)&$\pm 0.020$&(1.41)\\
&&&&&&&&&&&\\ 
HIP 116360  & 348.0437    & 56641.395 & 0.43503     &26.4670     &  359.664  & 20.362    &1.0271     & 87.734    & 14&  -1.147  &0.023\\
HD 221757  &$\pm 0.0054$ &$\pm 0.050$&$\pm 0.00036$&$\pm 0.0086$&$\pm 0.068$&$\pm 0.014$&$\pm0.0015$&$\pm 0.057$&(52)&  $\pm 0.048$  &(0.307)\\
        &             &           &             &            &           & 21.874    &0.9561     &94.251     & 14& 0.105     &0.025\\
        &             &           &             &            &           &$\pm 0.013$&$\pm0.0015$&$\pm 0.055$&(52)&$\pm 0.013$&(0.452)\\
 \hline
\label{tab:orbSB2}
\end{tabular}
\end{minipage}
\flushleft \bf
$^a$ The elements were derived fixing $P$ to the value obtained taking the measurements of \cite{Goldberg02} into account.
\end{table*}

\begin{figure*}
\includegraphics[clip=,height=20.5 cm]{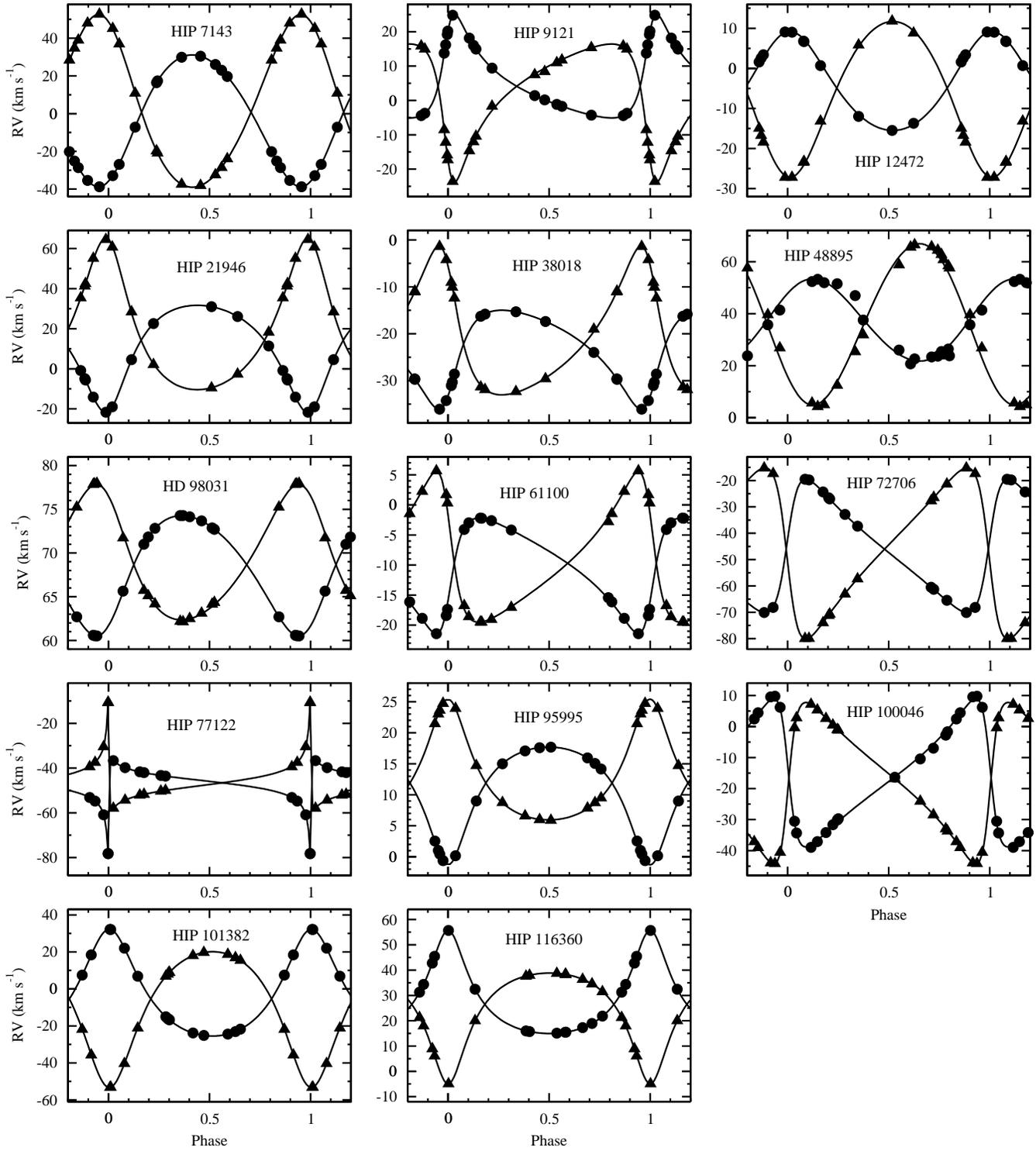} 
 \caption{The spectroscopic orbits of the 14 SB2; the circles refer to the SOPHIE RVs of the primary component, and the
triangles to the secondary. For each SB2, the RVs are shifted to the
zero point of the SOPHIE measurements of the primary component.}
\label{fig:orbSB2}
\end{figure*}

\begin{figure*}
\includegraphics[clip=,height=20.5 cm]{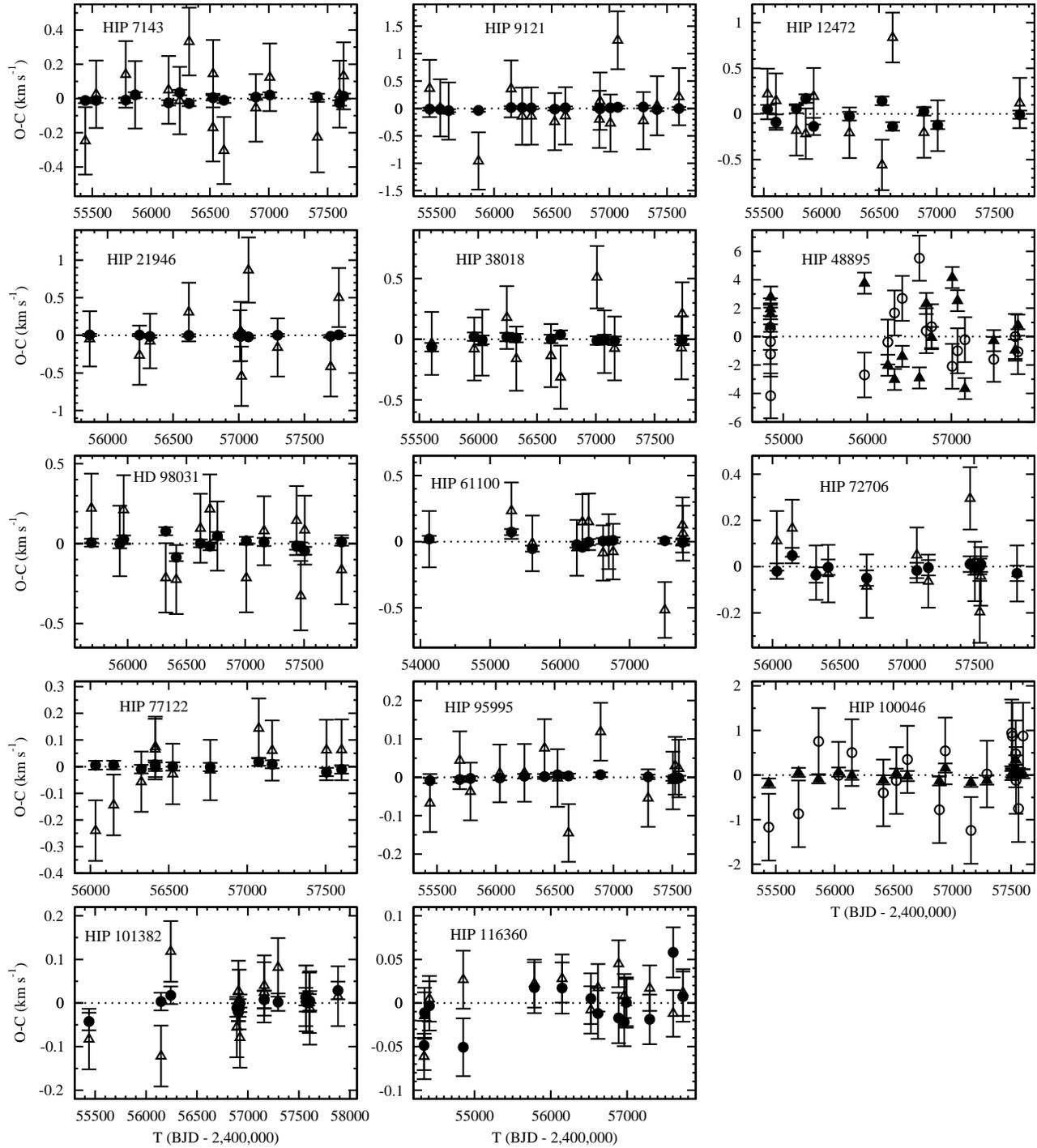} 
\caption{The residuals of the RVs obtained from {\sc todmor} for the 14 SB2s. The circles refer to the primary component, and the triangles to the secondary component. For readability, the residuals of the most accurate RV measurements are in filled symbols.}
\label{fig:resOrbSB2}
\end{figure*}


\section{Masses and parallaxes of HIP\,61100, HIP\,95995 and HIP\,101382}
\label{sec:masses}

When a visual orbit can be derived properly, an SB2 system with measured RV can be fully determined. Especially, the inclination can be evaluated and allows deriving the absolute mass of the system and of its components. Moreover with measurements independent of the Hipparcos 2 catalogue~\citep{vanLeeuwen07} it also allows verifying and correcting the Hippacos parallax taking into account the orbital motion. We found a visual orbit for 3 of the 14 SB2s presented in this paper; namely HIP\,61100, HIP\,95995 and HIP\,101382.

\subsection{HIP 61100} 
\begin{figure}
\includegraphics[clip=,height=5.4 in]{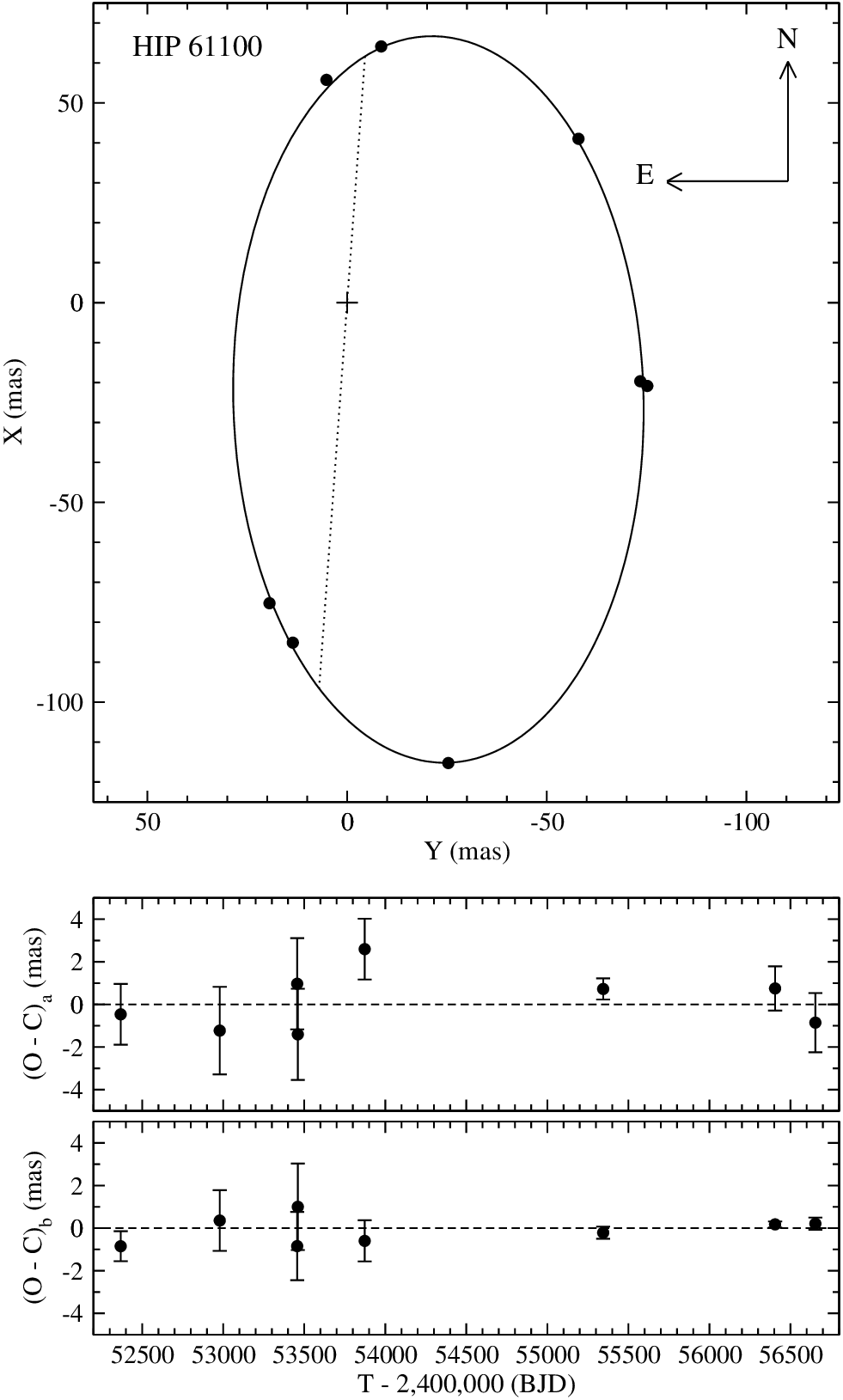}
 \caption{The visual part of the combined orbit of HIP 61100.Upper panel: the
visual orbit; the circles are the positions from Table~\ref{tab:HIP95995-interfero}; the node line is in dashes.
Middle panel: the residuals along the semi-major axis of the error ellipsoid. Lower panel: the residuals along the
semi-minor axis of the error ellipsoid.
}
\label{fig:HIP61100}
\end{figure}

Our RV measurements were combined to the speckle and interferometric observations used by
\cite{Schlieder2016}, which are summarized in Table~\ref{tab:HIP61100-interfero}. The uncertainties of the interferometric measurements were
corrected by 0.71 in order to obtain a visual orbit with a GOF $F_2=0$. These measurements are combined with our
RV measurements and led to the orbital elements given in Table~\ref{tab:VB-elements}.
The apparent orbit and its residuals are presented in Fig.~\ref{fig:HIP61100}.  
Our results are not really different from the preceding ones of \cite{Schlieder2016}, but slightly more accurate, with masses ${\mathcal M}_1$=(0.834$\pm$0.017)\,${\cal M}_\odot$  and ${\mathcal M}_2$=(0.640$\pm$0.011)\,${\cal M}_\odot$, improving the mass measurement accuracy for these stars by a factor of 2.5 compared to~\cite{Masda2016}.

Our estimation of the parallax in Table~\ref{tab:VB-elements}, is more accurate, but compatible, with that given
by the {\it Hipparcos 2} catalogue: $\varpi=(39.84 \pm 1.07)$~mas.
However, this value was derived ignoring the orbital motion. A correction of the Hipparcos parallax was derived from the elements in
Table~\ref{tab:VB-elements} and from the residuals of the Hipparcos astrometric solution. The new value
is then $\varpi=(40.75 \pm 1.24)$~mas, in reasonable agreement with our result.
No Tycho-Gaia Astrometric Solution \citep[{\it TGAS} hereafter; see Michalik, Lindegren \& Hobbs 2015;][]{GaiaDR1}
is available for this star, probably because of its orbital motion.

\begin{table}
 \caption{
The interferometric measurements of HIP 61100, taken from 
Schlieder~et~al.~(2016) and adapted to our purpose.
$X$ is oriented to North and $Y$ to East. $\sigma_a$ and $\sigma_b$ are the 
semi-major axis and the semi-minor axis of the ellipsoid error, respectively; they are corrected as explained in the text. 
$\theta_a$
is the position angle of the major axis of the ellipsoid error. $X$, $Y$ and $\theta_a$ are
for equinox 2000.
}
\begin{tabular}{@{}crrrrr@{}}
\hline
$T$-2,400,000 & $X$       & $Y$        & $\sigma_a$ & $\sigma_b$   & $\theta_a$     \\
(BJD)         & (mas)     & (mas)      & (mas)    & (mas)      & ($^{\rm o}$) \\
\hline
52367.9$^a$ &   41.031 &  -57.944 & 1.425  &   0.706 & 125.3  \\
52978.4 & -115.245 &  -25.349 & 2.053  &   1.425 & 102.4  \\
53456.7 &  -19.661 &  -73.413 & 2.138  &   1.604 &  75.0  \\
53460.6 &  -20.835 &  -75.166 & 2.138  &   2.031 &  74.5  \\
53872.8 &   55.760 &    5.181 & 1.425  &   0.969 &   5.3  \\
55344.1 &  -75.230 &   19.436 & 0.499  &   0.285 & 165.5  \\
56405.2 &   64.134 &   -8.537 & 1.041  &   0.143 &  82.4  \\
56653.2$^a$ &  -85.120 &   13.605 & 1.390  &   0.285 &  80.9 \\
\hline
\end{tabular}
\label{tab:HIP61100-interfero}
\flushleft 
$^a$ We have merged two measurements performed at the same epoch.
\end{table}

\subsection{HIP\,95995} 
\begin{figure}
\includegraphics[clip=,height=5.4 in]{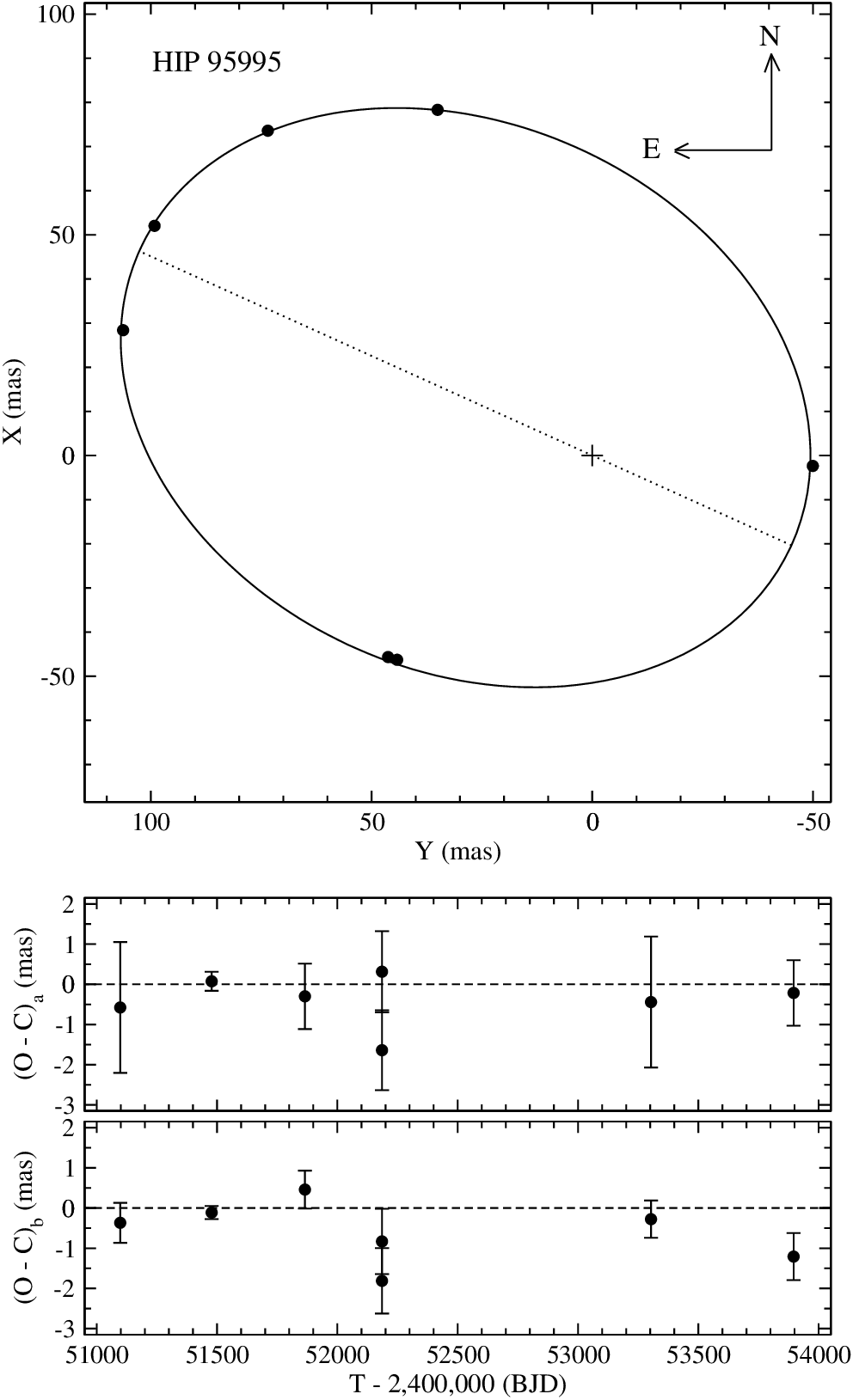}
 \caption{The visual part of the combined orbit of HIP 95995.Upper panel: the
visual orbit; the circles are the positions from Table~\ref{tab:HIP95995-interfero}; the node line is in dashes.
Middle panel: the residuals along the semi-major axis of the error ellipsoid. Lower panel: the residuals along the
semi-minor axis of the error ellipsoid.
}
\label{fig:HIP95995}
\end{figure}

This star is the close visual binary MCA~56. \cite{Masda2016} combined RV measurements and the interferometric
measurements provided by the Fourth Catalog of Interferometric Measurements of Binary Stars\footnote{http://www.usno.navy.mil/USNO/astrometry/optical-IR-prod/wds/int4} 
\citep[Third catalogue:][]{Hartkopf01}
to derive the masses of the components: ${\cal M}_1$=(0.89$\pm$0.08)\,${\cal M}_\odot$ 
and ${\cal M}_2$=(0.83$\pm$0.07)\,${\cal M}_\odot$. We found that the less accurate measurements in the
``Fourth Catalog'' were also the less reliable ones, since their errors are much larger than their 
uncertainties; therefore, we derived the visual orbit taking into account only the measurements with uncertainties
smaller than 2 mas. These measurements are presented in Table~\ref{tab:HIP95995-interfero}. We applied to
these uncertainties a correcting factor of 0.81 in order to get the apparent orbit with the GOF $F_2=0$. The combination of the relative positions with our RVs leads to the parameters in Table~\ref{tab:VB-elements}. The apparent orbit and its residuals are presented in Fig.~\ref{fig:HIP95995}. We found the masses ${\cal M}_1$=(0.833$\pm$0.031)\,${\cal M}_\odot$ and ${\cal M}_2$=(0.812$\pm$0.030)\,${\cal M}_\odot$

We found the parallax $\varpi=(56.10 \pm 0.81)$~mas, which is slightly different
from that given in the {\it Hipparcos 2} catalogue: $\varpi=(58.96 \pm 0.65)$~mas. 
This is due to the orbital motion with a period close to one year:
Correcting the Hipparcos parallax for
this motion leads to $\varpi=(57.15 \pm 0.31)$~mas, in acceptable agreement with our result.
The parallax from {\it TGAS} \citep{Michalik2015,GaiaDR1} is $\varpi=(58.37 \pm 0.54)$~mas; the difference 
probably comes from the fact that the orbital motion was ignored in the calculation of {\it TGAS}.

\begin{table*}
 \caption{The combined VB+SB2 elements of HIP 61100, HIP 95995 and HIP 101382. For HIP 61100 and HIP 95995, the elements
are derived from a combined VB+SB2 solution.  
For consistency with the SB orbits and with the
forthcoming astrometric orbit, $\omega$ refer to the motion of the primary component.}
\begin{tabular}{@{}lccc@{}}
\hline
  &  HIP\,61100 & HIP\,95995 & HIP\,101382\\
\hline
$P$ (days)              &   1285.31 $\pm$ 0.27    & 494.307 $\pm$  0.012   & 57.32176 $\pm$ 0.00010$^a$     \\
$T_0$ (BJD-2400000)     &  56492.13 $\pm$ 0.35    & 56549.505 $\pm$ 0.043  & 56627.3786 $\pm$ 0.0075$^a$    \\
$e$                     &  0.51130 $\pm$  0.00093 & 0.38933 $\pm$ 0.00029  & 0.43503  $\pm$ 0.00036$^a$     \\
$V_0$ (km~s$^{-1}$)     &  -9.7113 $\pm$ 0.0096   &  11.8932 $\pm$ 0.0018  & -5.4108  $\pm$ 0.0057$^a$      \\
$\omega_1$ ($^{\rm o}$) &   244.50 $\pm$ 0.16     & 180.325 $\pm$ 0.041    & 357.195  $\pm$ 0.064$^a$       \\
$\Omega$($^{\rm o}$; eq. 2000) &  355.83 $\pm$ 0.24   & 245.72 $\pm$ 0.13  & 334.960  $\pm$ 0.070$^b$       \\
$i$  ($^{\rm o}$)       &   58.63 $\pm$  0.46     &  146.15 $\pm$ 0.46     & 99.364   $\pm$ 0.080$^b$       \\
$a$ (mas)               &      102.19             &    81.03               & 15.378   $\pm$ 0.027$^b$       \\
${\cal M}_1$ (${\cal M}_\odot$) &  0.834 $\pm$ 0.017 & 0.833 $\pm$ 0.031   & 0.8420   $\pm$ 0.0014      \\
${\cal M}_2$ (${\cal M}_\odot$) & 0.640 $\pm$ 0.011  & 0.812 $\pm$ 0.030   & 0.66201  $\pm$ 0.00076     \\
$\varpi$ (mas)          &    38.82 $\pm$  0.23    & 56.10 $\pm$ 0.81       & 46.131   $\pm$ 0.084       \\
$d_{2-1}$ (km~s$^{-1}$) &   0.097 $\pm$  0.064    & 0.112 $\pm$ 0.021      & 0.187    $\pm$ 0.020$^a$       \\
$\sigma_{(o-c)\;VB}$ (mas)  &    1.04             &  0.673                 &          -                 \\
$\sigma_{(o-c)\;RV}$ (km~s$^{-1}$)  & 0.019, 0.202   &  0.0049, 0.066      &  0.017, 0.063$^a$              \\
\hline \\
$M_{A}$  (mag)        &    4.06 $\pm$ 0.25$^d$      &   5.98 $\pm$ 0.02$^e$      &     $\ldots$   \\
$M_{B}$  (mag)        &    4.77 $\pm$ 0.27$^d$      &   6.24 $\pm$ 0.03$^e$      &     $\ldots$   \\
$Y$                     &       $0.245-0.27$      &     $0.245 - 0.279$    &     $\ldots$   \\
age    (Gyr)            &        0.4$^c$          &      $2.2 - 7.9$       &     $\ldots$ \\
${\rm [Fe/H]}$ (dex)&        -0.17            &      -0.33$~-~$-0.17 &     $\ldots$ \\
\hline
\end{tabular}
\label{tab:VB-elements}
\flushleft 
$^a$ From Table~\ref{tab:orbSB2}.\\
$^b$ From the ``Full Fit'' solution of \cite{Torres2002}. \\
$^c$ Age of the UMA Group \citep{Jones2015} \\
$^d$ K-band magnitude \\
$^e$ V-band magnitude \\
\end{table*}

\begin{table}
 \caption{Same as Table~\ref{tab:HIP61100-interfero}, but for HIP 95995.
Only the positions more accurate than 2~mas 
are selected from the Fourth Catalogue of
Interferometric Measurements of Binary Stars. 
}
\begin{tabular}{@{}crrrrr@{}}
\hline
$T$-2,400,000 & $X$       & $Y$        & $\sigma_a$ & $\sigma_b$   & $\theta_a$     \\
(BJD)         & (mas)     & (mas)      & (mas)    & (mas)      & ($^{\rm o}$) \\
\hline
51097.8     &  -2.366 &  -49.944 & 1.628  & 0.497 &  87.3 \\
51478.0     &  78.322 &   35.032 & 0.236  & 0.163 & 114.1 \\
51865.3     &  52.047 &   99.172 & 0.814  & 0.472 &  62.3 \\
52185.5     & -46.283 &   44.203 & 0.993  & 0.814 &  46.3 \\
52185.5     & -45.654 &   46.268 & 1.009  & 0.814 &  44.6 \\
53303.2     &  28.382 &  106.275 & 1.628  & 0.464 &  75.0 \\
53896.0     &  73.586 &   73.492 & 0.814  & 0.586 &  44.9 \\
\hline
\end{tabular}
\label{tab:HIP95995-interfero}
\end{table}

\subsection{HIP 101382} 

For this system, \cite{Torres2002} already derived the orbital elements from the combination of observations made at
the Palomar Testbed Interferometer, with RV measurements. Unfortunately, they did not provide the positions of the
secondary component with respect to the primary, so we cannot compute a combined orbit as for the two preceding binaries.
However, comparing the elements of their ``full fit''
with those derived from our RVs, it appears that, expressed in unit of uncertainties, the discrepancies for period, 
periastron epoch, eccentricity and periastron longitude are -0.41, 0.91, 1.81 and 0.69
respectively. These values are all between -2 and +2, indicating  a nice agreement between their
elements and our solutions. With the inclination derived from their ''full fit'', we found the new masses:
${\cal M}_1$=(0.8420$\pm$0.0014)\,${\cal M}_\odot$ and ${\cal M}_2$=(0.66201$\pm$0.00076)\,${\cal M}_\odot$, improving the accuracy on these masses by a factor of 10 compared to~\cite{Torres2002}. 

Moreover, we evaluated a new measurement of the parallax $\varpi=(46.131 \pm 0.084)$~mas. For comparison, the {\it Hipparcos 2} catalogue gives $(45.35 \pm 0.43)$~mas, which becomes $(45.31 \pm 0.44)$~mas when our orbital elements are applied to the residuals of the Hipparcos astrometric
solution. The {\it Hipparcos 2} parallax is thus marginally compatible with ours, although slightly underestimated.
The parallax from {\it TGAS} \citep{Michalik2015,GaiaDR1} is $\varpi=(46.61 \pm 0.83)$~mas, in good agreement 
with our result, although the orbital motion was not taken into account in the calculation.

\section{Initial stellar parameters of HIP\,61100 and HIP\,95995}
\label{sec:models}

Having derived very accurate masses for HIP\,61100 and HIP\,95995 allowed us to characterize the two components of these binaries in terms of initial helium content and age. For that purpose, we modeled the two components following the stellar model optimisation method described in~\cite{Lebreton2014}. We adopted the reference set of stellar input physics described in that paper and the {\verb+Cesam2k+} stellar evolution code~\citep{Morel2008}. The observational constraints considered for the models are the masses of the two components herebefore determined, their effective temperatures and luminosities, and the present metallicity of the primary component. We point out that we decided not to model HIP\,101382 because this binary system is enriched in $\alpha$-elements, with [$\alpha\text{/Fe}$]=$0.36$~\citep{Torres2002}. As discussed by \cite{Torres2002}, a proper modeling would require to calculate new opacity tables which is beyond the scope of the present paper.

In support of the previous estimation of stellar parameters given in Table~\ref{tab:stellpar}, we used the code \verb+iSpec+~\citep{Blanco2014} to verify the primary stellar parameters. Results for HIP\,61100 and HIP\,95995 are discussed below and appended to Table~\ref{tab:VB-elements}.

\subsection{HIP\,61100}

To derive the luminosities of the components, we proceeded as follows. First, we used the system K band magnitude, $K$=5.662$\pm$0.020 from 2MASS~\citep{Cutri2003}, the magnitude difference of the two components in the K band, $\Delta K$=0.71$\pm$0.02 \citep{Schlieder2014}, and the spectroscopic parallax derived in the present study. We obtained the absolute magnitudes $M_{K, A}$=4.06$\pm$0.25 and
$M_{K, B}$=4.77$\pm$0.27\,mag. Then, in the calculation of the stellar models, we derived the luminosity using the bolometric corrections $BC_K(T_\mathrm{eff},\log g,\mathrm{[Fe/H]})$ of~\cite{Casagrande2014}.  

The \verb+iSpec+ derivation gave $T_\text{eff}$=5049$\pm$43\,K, $\log g$=4.52$\pm$0.15\,dex, and [Fe/H]=-0.13$\pm$0.10, as consistent with what derived in Table~\ref{tab:stellpar}. Therefore, we constrained the stellar models with the effective temperatures of Table~\ref{tab:stellpar}. The choice of the metallicity is more delicate. We therefore considered three possible values of the metallicity ($\mathrm{[Fe/H]}$=-0.13, -0.18, -0.33\,dex) covering the range reported in the literature displayed in the SIMBAD database (Wenger et al. 2000)..

We further assumed that the stars have a common origin and therefore share the same initial metallicity, helium abundance and age.
Their initial helium abundance in mass fraction should be higher than the primordial value $Y_p$\,\!$\sim$0.245 \citep[see e.g.][]{Peimbert2016,Izotov2014}. Furthermore, according to~\cite{King2003}, the system is a bona-fide member of the UMA Group nucleus. Therefore, we assumed that the common age of the components is 400\,Myr, i.e. the age of the UMa group~\citep{Jones2015}.

\begin{figure}
\includegraphics[clip=,width=\columnwidth]{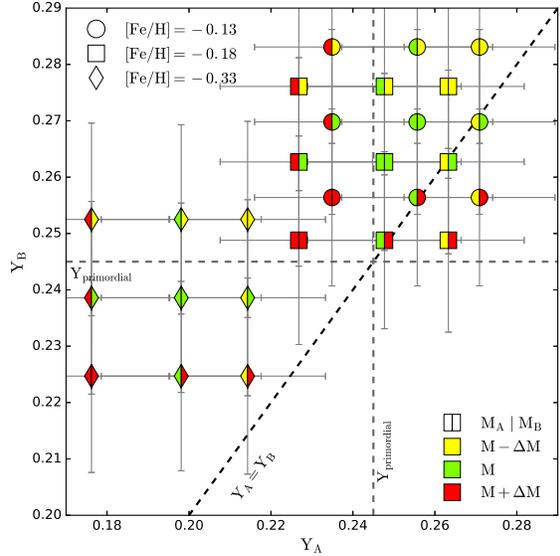}
\caption{\label{fig:He_abun} Initial He abundance of the primary and secondary components of HIP\,61100 inferred from different sets of optimized models. Different symbols correspond to different assumptions on the measured metallicity [Fe/H], as found for the same star HIP\,61100A in the literature. Independent \texttt{iSpec} derivation of HIP\,61100 metallicity gives [Fe/H]=-0.13$\pm$0.10 dex. Each symbol is separated in two, primary star on the left side and secondary on the right. Each mass is varied on its confidence interval, leading to different estimation of the He abundance.}
\end{figure}

The model optimisation provides the initial helium abundance for each star. The results are shown in Fig.~\ref{fig:He_abun}.

We first note that low values of the metallicity  ($\mathrm{[Fe/H]}$=$-0.33$\,dex) can be excluded because they would lead to a sub-primordial initial helium abundance of the system. On the other hand, considering also the constraint that the two components have the same initial helium abundance, we find good compatibility with models on the higher metallicity case, with $\mathrm{[Fe/H]} \sim -0.15$, provided that the primary mass is on the lower bound and secondary mass on the upper bound of their confidence interval. Finally, we also explored the possibility that the stars have an age of 500\,Myr, as assumed by~\cite{Schlieder2016} but did not find any satisfactory solution with the subsolar $\mathrm{[Fe/H]}$ values considered here.

\subsection{HIP\,95995}

To derive the luminosities, we took the parallax derived here, the system V band magnitude ($V$=6.607$\pm$0.010) from Tycho~2  \citep{Hog2000}, and the magnitude difference of the two components in the $V$ band ($\Delta V$=0.26$\pm$0.03) which we calculated as the mean of interferometric values listed in Table 2 of~\cite{Masda2016}, but keeping the values with given error bars only. We obtained the absolute magnitudes $M_{V, A}$=5.98$\pm$0.02 and $M_{V, B}$=6.24$\pm$0.03\,mag. Then,  we applied the bolometric corrections of~\cite{Casagrande2014}. 
We did not include extinction, since it is usually expected to be very small for a star at less than 20 pc.

The primary stellar parameters derived from \verb+iSpec+ are $T_\mathrm{eff}$=4972$\pm$32\,K and $\mathrm{ [Fe/H]}$=-0.45$\pm$0.27,
in reasonable agreement with those derived in Table 2. Therefore, we constrained the stellar models with the effective temperatures of Table~\ref{tab:stellpar}. 
Since \cite{Casagrande2011} rather derived 
$\mathrm{ [Fe/H]}$=-0.17,  we performed two sets of models, one with the metallicity determined here ($\mathrm{ [Fe/H]}$=-0.33)  and one with $\mathrm{ [Fe/H]}$=-0.17.

In the case of HIP 95995, we do not have constraints on the age. However, the star is classified as inactive by \cite{Gray2003} which is not in favor of young ages. Moreover,~\cite{Casagrande2011} gives a rough estimation of the age of this system, 13.8$\pm$6\,Gyr. 

To model the system, we first optimized models of the primary component, adjusting the age, initial helium abundance and metallicity; and then we searched for a model of the secondary component by fixing its age and initial composition equal to those of the primary. 
Unfortunately, despite the improvement on the mass, the stellar model of HIP\,95995 remains rather poorly constrained, due to possible misestimation of the secondary's stellar parameters. We thus eventually discarded the secondary's constraint and only considered the contribution of the primary in the following. 

No acceptable solution (Y$>$Y$_\text{primordial}$ and age$>$1\,Gyr) is found for the upper part of the mass confidence interval (${\cal M}_A$\,\!$>$0.833\,${\cal M}_\odot$), and decreasing or increasing the metallicity still leads to reject the models. At ${\cal M}_A$=0.833\,${\cal M}_\odot$, the best models stand around $\mathrm{ [Fe/H]}$=-0.17; they lead to an age range $\sim$2.2-5.3\,Gyr and initial helium abundance $Y_0$=0.250-0.265. Finally, for ${\cal M}_A$\,\!$<$0.833\,${\cal M}_\odot$, the most suitable models have ages in the range 2.4-7.9 Gyr, initial helium abundance in the range 0.245-0.279, and metallicity between -0.17 and -0.33.

\section{Summary and conclusion}
\label{sec:conclusion}

Thanks to new SOPHIE spectra of 14 SB2s, four of which are newly identified SB2 (paper I), and the use of the {\sc todmor} code, we derived new better accurate orbital solutions to the RV measurements of these binaries. The projected masses ${\cal M}\sin^3 i$ were calculated for all 14 SB2s, with an average accuracy of 1.0$\pm$0.2\,\%, with extreme cases such as the rapid rotator HIP\,48898 with $\sigma({\cal M}\sin^3 i)$\,\!$\sim$4\,\%, or HIP\,101382 with $\sigma({\cal M}\sin^3 i)$\,\!$\sim$0.12\,\%. 

For HIP\,61100, HIP\,95995 and HIP\,101382, archival interferometric measurements allowed us to fully constraint the systems and derive masses for components (A,B) with accuracies respectively (2.0, 1.7)\,\%, (3.7, 3.7)\,\%, and (0.2, 0.1)\,\%. The stellar evolution code \verb+Cesam2k+~\citep{Morel2008} applied to HIP\,61100 and HIP\,95995, led to constrain their age, metallicity and initial helium content. HIP\,61100 was found slightly overabundant in He with respect to primordial helium abundance, with $\mathrm{ [Fe/H]}\sim-0.15$\,dex and an age close to 400\,Myr, while HIP\,95995 was harder to constrain, assuming a relatively old star with age$>$1\,Gyr, and using only primary star's mass and stellar parameters, led to a possible overabundance in He, with -0.33$<$[Fe/H]$<$0.17. 

Although we could not calculate stellar evolution models of HIP\,101382, the masses of the SB2 components that we derived reached the level of 0.1\,\% accuracy. In the future, this star will likely become a reference for validating masses derived from {\it Gaia}.  

Added to the systems already published in papers II and III, we have now 6 binaries observed with SOPHIE and interferometric instruments which may be used to verify the masses that will be derived from {\it Gaia}. This number will continue to increase until the completion of the programme.

\section*{Acknowledgments}

We sincerely thank the anonymous reviewer for his careful reading of our manuscript and valuable comments. This project was supported by the french INSU-CNRS ``Programme National de Physique Stellaire'', ``Action Sp\'{e}cifique {\it Gaia}'', and the Centre National des
Etudes Spatiales (CNES). We are grateful to the staff of the
Haute--Provence Observatory, and especially to Dr F. Bouchy, Dr H. Le Coroller, Dr M. V\'{e}ron, and the night assistants, for their
kind assistance. We made use of the SIMBAD database, operated at CDS, Strasbourg, France. This research has received funding from the European Community's Seventh Framework Programme (FP7/2007-2013) under grant-agreement numbers 291352 (ERC)

\label{lastpage}

\bsp

\end{document}